\documentclass{mn2e}
\usepackage{epsfig}

\newcommand{\mnref}[1]{\hangindent=0.5in \hangafter=1 #1 \par}
\newenvironment{refs}{\parindent=0pt}{\parindent=1.5em}
\newcommand{\mn}{MNRAS}

\newcommand{\aj}{AJ}
\newcommand{\apj}{ApJ}
\newcommand{\apjs}{ApJS}
\newcommand{\aaa}{A\&A}
\newcommand{\aas}{A\&AS}
\newcommand{\mic}{\mbox{$\mu{\rm m}$}}
\newcommand{\smpy}{\mbox{M$_{\odot}$ yr$^{-1}$}}

\newcommand{\Msolar}{\mbox{\,$\rm M_{\odot}$}}
\newcommand{\Lsolar}{\mbox{\,$\rm L_{\odot}$}}
\def\gs{\mathrel{\raise1.16pt\hbox{$>$}\kern-7.0pt
\lower3.06pt\hbox{{$\scriptstyle \sim$}}}}
\def\ls{\mathrel{\raise1.16pt\hbox{$<$}\kern-7.0pt
\lower3.06pt\hbox{{$\scriptstyle \sim$}}}}

\newcommand{\gtappeq}{\raisebox{-0.6ex}{$\,\stackrel
{\raisebox{-.2ex}{$\textstyle >$}}{\sim}\,$}}

\title[The Population of the Galactic Plane as Seen by MSX]
{The Population of the Galactic Plane as Seen by MSX}
\author[S.L. Lumsden, M.G. Hoare, R.D. Oudmaijer and D. Richards]
{S.L. Lumsden, M.G. Hoare, R.D. Oudmaijer and
D. Richards\\ {} {\em Department of Physics and Astronomy, University
of Leeds, Leeds LS2 9JT, UK}\\ {Email -- sll@ast.leeds.ac.uk,
mgh@ast.leeds.ac.uk, roud@ast.leeds.ac.uk}\\ }
\begin{document}

\label{firstpage}

\maketitle

\begin{abstract}
The combination of mid-infrared data from the MSX satellite mission
and ground-based near-infrared photometry is used to characterise the
properties of the mid-infrared population of the Galactic plane.  The
colours of the youngest sources still heavily embedded within their
natal molecular clouds are in general different from evolved stars
shrouded within their own dust shells. Our main motivation is to use
MSX for an unbiased search for a large ($\sim$ 1000) sample of massive
young stellar objects (MYSOs). A simple analysis shows that the MSX
point source catalogue should contain most of the MYSOs within our
Galaxy. We develop colour selection criteria using combined near- and
mid-infrared data for MYSOs, which produces a list of 3071 objects,
excluding the galactic centre region.  The programme of follow-up
observations already underway to separate the MYSOs from compact H~II
regions and other remaining objects is briefly described. We also show
that these data can be used, just as IRAS data has been previously, to
provide a separation between evolved stars with carbon rich and oxygen
rich dust.  These data may also be used to search for evidence of dust
around normal main sequence stars, such as low mass pre-main
sequence stars or the Vega-excess class of objects where debris disks
are presumed to remain from the planet formation process.  We discuss
the accuracy and completeness of the MSX point source catalogue, and
show that the errors present tend to be of a kind that is not
significant for the main stellar populations we discuss in this paper.

\end{abstract}


\section{Introduction}

Objects embedded in dust play an important role throughout
stellar evolution. Stars are born out of dusty molecular clouds whilst in
the late stages of evolution they will invariably go through phases where
they generate their own dust during heavy mass-loss. These objects
emit most prodigiously in the infrared region of the spectrum with
dust temperatures usually in the range of 30-300 K. To study these
objects there is an undeniable need for unbiased samples and these
usually have to originate from the infrared where most of their
bolometric luminosity emerges. 

Unbiased infrared surveys started with the rocket-borne AFGL mid-infrared
survey of Price (1977) and then the first far-infrared satellite survey came
with the IRAS mission. Many studies have been made using IRAS colours to select
and classify objects of a variety of types (see Beichman 1987).  These can be
split roughly into studies of evolved stars and studies of young stars.  In the
former category, Van der Veen \& Habing (1988) and Walker \& Cohen (1988) used
12, 25 and 60$\mu$m colours to classify the stars in the IRAS point source
catalogue (IRAS PSC). These were mostly Asymptotic Giant Branch (AGB) stars
that could be separated into those that were either oxygen-rich or carbon-rich,
although first ascent Red Giant Branch (RGB) stars also have dust excesses
(Plets et al.\ 1997; Jura 1999). Searches for evolved stars where mass-loss had
stopped and the dust shells have become detached were made by Oudmaijer et
al.\ (1992) for post-AGB stars and for planetary nebulae (e.g. Pottasch et
al.\ 1988).  Another example of the potential of IRAS for stellar astronomy was
the discovery that some main sequence stars showed excess dust emission due to
a remnant disk, now known as Vega-type or Vega-excess stars (Aumann 1985).

The other major area that benefited from IRAS was the study of the young
stellar population in our galaxy.  IRAS colour-colour diagrams were used by
Hughes \& MacLeod (1988) to separate H~II regions from planetary nebulae and
reflection nebulae for example. Prusti, Adorf \& Meurs (1992) used more
sophisticated techniques to extract low-mass young stellar object candidates
from the IRAS PSC.  Our main motivation is to build a large unbiased sample of
massive young stellar objects (MYSOs), also known after the prototype in Orion
as BN objects (see Henning et al.\ 1984). By this we mean a luminous
($\gtappeq10^{4}$ \Lsolar\ or early B star), embedded source that has not yet
reached the stage when the emergent Lyman continuum is sufficient to ionize the
surrounding interstellar medium to form an H~II region. Due to the short
contraction timescale compared to the collapse timescale it is thought that
these objects are already core H-burning. However, it is also likely that
accretion is still ongoing. They usually possess strong ionized stellar winds
(e.g. Bunn, Hoare \& Drew 1995) and drive powerful bipolar molecular outflows
(Lada 1985). The 30 or so currently well-known MYSOs form a heterogeneous
sample (Wynn-Williams 1982; Henning et al.\ 1984), often found serendipitously,
and may not be representative of the class as a whole.  There is an obvious
need for a much larger well-selected sample.

It is reasonable at this point to ask how many MYSOs are in the entire galaxy.
We can make a crude estimate by normalizing a stellar mass formation rate over
a Salpeter IMF to yield the currently accepted star formation rate in the
galaxy of about 6\smpy\ (G\"{u}sten \& Mezger 1982). We then simply used the
Kelvin-Helmholtz timescale for contraction to the main sequence using
mass-radius and mass-luminosity relations for main sequence stars.  Although
not strictly correct this does give appropriate timescales for the YSO phases
that we are interested in, namely about 10$^{4}$ years for the most massive O
star and 10$^{5}$ years for the intermediate mass Herbig Be type stars. This
predicts about 1000 massive (15-100\Msolar) YSOs in the galaxy, which is well
over an order of magnitude larger than the current sample. A sample of this
size is required when studying the properties of MYSOs as functions of mass,
age, metallicity, etc. Hence, the task in reality is close to that of finding
every MYSO in the galaxy.

Previous attempts at systematic searches for MYSOs have been made using IRAS
data.  Campbell, Persson \& Matthews (1989) in a series of papers followed-up
400 colour-selected IRAS sources that were also bright, unconfused and not
identified with a known source. The problem with any infrared colour-selection
procedure is that there are other types of source that have very similar
infrared colours. Basically any heat source inside an optically thick dust
cloud will produce an emergent spectral energy distribution that depends mostly
on the optical depth of the cloud.  The population with most similar infrared
colours to MYSOs are compact H~II regions, for which they are the progenitors.
Both are are still deeply embedded in dense molecular clouds. Young compact,
planetary nebulae and very dusty evolved stars can also have similar infrared
colours. Campbell et al.\ used single-element large aperture (8~arcsecond)
near-infrared photometry to observe their IRAS candidates. They found a total
of 115 candidate YSOs that were bright and red near-infrared sources, with
excess emission in the $K$-band presumed to be from hot circumstellar dust.
These candidates have luminosities in the range 10-10$^4$\Lsolar\ however so
only just meet our criteria for a genuinely massive YSO.  Recent spectroscopy
presented by Ishii et al.\ (2001) suggests that some are not YSOs since the HI
Br$\gamma$ equivalent width is much closer to that seen in H~II regions (eg
Lumsden \& Puxley 1996) or PN (eg Lumsden, Puxley \& Hoare 2001), than massive
YSOs (eg Porter, Drew \& Lumsden 1998).  This highlights the importance of
obtaining sufficient follow up data to confirm the nature of any candidate
MYSO.

Chan, Henning \& Schreyer (1996) used similar selection criteria to Campbell et
al.\ and derived a list of 254 MYSO candidates. However, no attempt was made to
eliminate compact H~II regions from the sample. Indeed over 100 of the objects
are listed as being strong radio sources, much brighter than the weak stellar
wind emission from nearby MYSOs (S$_{\nu}<$10 mJy) (e.g. Henning, Pfau \&
Altenhoff 1990; Tofani et al.\ 1995), and are therefore very likely to be H~II
regions. Palla et al.\ (1991) put together a colour-selected bright IRAS sample
of 260 objects that avoided known H~II regions from single-dish radio surveys.

Of course, any infrared colour-selected sample of compact H~II regions
by the converse will also contain large numbers of genuine MYSOs. Wood
\& Churchwell (1989a) developed 12, 25 and 60\mic\ colour-selection
criteria based on ultra-compact H~II regions detected in their VLA
survey. They applied this to the whole IRAS PSC and found 1646
embedded massive stars. Codella, Felli \& Natale (1994) found that
much more evolved, diffuse H~II regions as well as compact H~II
regions satisfy the Wood \& Churchwell criteria and Ramesh \&
Shridharan (1997) found evidence for significant contamination of the
Wood \& Churchwell sample by cores heated by less massive stars.
Kurtz, Churchwell \& Wood (1994) carried out a VLA survey of 59 bright
IRAS sources satisfying the Wood \& Churchwell colour criteria. They
found that 80\% had compact radio emission and are thus compact H~II
regions, although some of these are weak enough to be MYSOs. Indeed
several well-known MYSOs were recovered in their study. Sridharan et
al.\ (2002) selected a sample of 69 bright, northern MYSOs that
satisfied the Wood \& Churchwell criteria as well as having dense gas
traced via CS emission and undetected in single-dish radio continuum
surveys. The latter condition was to avoid complexes and H~II regions,
although VLA follow-ups showed that nearly one third of their sample
were still compact H~II regions. Their sample was deliberately biased
towards finding isolated MYSOs for detailed high resolution studies.

Walsh et al.\ (1998) carried out high resolution radio observations of IRAS
sources originally selected in the same way as Wood \& Churchwell, but mostly
known to emit either radio continuum or methanol maser emission from previous
single dish observations. Methanol masers have been known to trace massive star
forming regions for some time (Menten 1991; Caswell et al.\ 1995). High
fractions of IRAS selected sources and well known MYSOs are maser sources, not
only methanol, but H$_{2}$O and OH as well. However, unfortunately any one
masing species cannot be guaranteed to be present in every MYSO. Indeed a
recent search for methanol emission in a sample of well-known MYSOs found that
none of them were detected (Gibb, Hoare \& Minier, in preparation). So although
there are great advantages in using strong, unobscured radio line emission
techniques for galactic searches they will not be unbiased. Similarly, radio
continuum emission from MYSOs is simply too weak to allow continuum mapping to
be used in providing an unobscured search method for these objects.

Clearly therefore any unbiased search must start from infrared
surveys. Preferably these should be at far-infrared wavelengths where the
emission from the regions peaks and the extinction across the galaxy is
negligible. Purely near-infrared surveys would be biased towards nearby and
less heavily embedded (more evolved) sources due to the high extinction in the
near-infrared. However, the concentrations of massive stars within
$\sim\pm1^\circ$ of the Galactic plane (Reed 2000; Becker et al.\ 1994) means
there is considerable source confusion in the IRAS survey due to its large beam
(45~arcseconds $\times$ 240~arcseconds at 12\micron).  Recently the MSX
satellite carried out a much higher spatial resolution survey (18~arcseconds)
of the Galactic plane ($|b|<5^\circ$) at 8, 12, 14 and 21$\mu$m, completely
superseding IRAS. This provides the starting point for our unbiased search for
MYSOs.  The aim of this paper is to demonstrate that this mid-infrared data
does provide a valuable means of locating these MYSOs, and to show how the
various Galactic populations that emit in the mid-infrared can be characterised
from such data.

\section{Stellar populations in the mid-infrared}

\subsection{The MSX Point Source Catalogue}

The Mid-course Space Experiment (MSX) satellite mission included an
astronomy experiment (SPIRIT III) designed to acquire mid-infrared
photometry of sources in the Galactic Plane as well as other regions
either not covered by IRAS, or where the IRAS data were confused due
to the high source density.  MSX has a raw resolution of 18.3$''$, a
beam size 50 times smaller than IRAS at 12 and 25$\mu$m,

The MSX satellite observed 6 bands between 4 and 21$\mu$m.  Full details of the
mission can be found in Price et al.\ (2001).  The most sensitive data, by a
factor of $\sim10$, were acquired using an $8.3\mu$m filter (band A: bandwidth
3.4$\mu$m).  The point source sensitivity was similar to that of the IRAS
12$\mu$m band, at about 0.1 Jy.  Data obtained using both the 14.7 and
21.3$\mu$m filters (bands D and E: widths 2.2 and 6.2$\mu$m respectively),
although less sensitive, were also of value to us since we are primarily
interested in red objects. The E band can be compared to the IRAS 25$\mu$m band
(see Egan et al.\ 1999).  The two filters at 4$\mu$m were not designed
primarily for astronomical observations and little useful data is obtained from
them.  The final filter at 12.1$\mu$m (band C: width 1.7$\mu$m) is less
sensitive than band D and consequently of lesser value, though we did use it
where appropriate.  Band C is effectively a narrower version of the IRAS
12$\mu$m filter (again see Egan et al.\ 1999).

The vast majority of the 300~000 point sources detected by MSX are found only
as 8$\mu$m detections and are not found in the IRAS Point Source Catalogue.
The full properties of the sources found in the MSX Point Source Catalogue
(hereafter MSX PSC) are outlined in Egan et al.\ (1999).  We used v1.2 of the
catalogue.  Since colour information is needed to begin to classify objects, we
have mainly used a subset of the full catalogue, consisting of those sources
with quality flags in the MSX PSC of 2 or greater in the A, D and E bands.
Egan et al.\ (1999) give the definition of sources with a quality of 2 as fair
or poor.  Their Table 5 gives the actual definition in terms of signal-to-noise
and number of repeat sightings from the separate scans of the Plane.  Our
`multi-colour' subset comprised 14897 sources.  Hereafter in this paper we will
refer to these MSX bands by their central wavelengths as the 8, 12, 14 and
21$\mu$m bands.

We cross-correlated the multi-colour MSX sample with the IRAS Point Source
Catalogue.  The IRAS PSC contains 9303 (62\%) of the subset of MSX sources.
Virtually all of the matches had detections in the short wavelength bands as
might be expected (296 were missing at 12$\mu$m, 218 at 25$\mu$m and only 20 at
both 12 and 25$\mu$m).  However, 145 were upper limits at 60$\mu$m, 2794 were
upper limits at 100$\mu$m and 5373 were upper limits in both these bands.  The
results are consistent with the increased source confusion in the larger IRAS
beam, especially at long wavelength.  A crude comparison of the measured fluxes
in the IRAS PSC and MSX PSC for objects detected in both is also consistent
with this hypothesis.  Crucially only 25\% have reliable data at 60$\mu$m which
is a key ingredient in previous source classification using IRAS data.  This
highlights the difficulties present in using IRAS data to detect and classify
sources within the inner regions of the Galactic Plane.

We checked the claimed astrometric accuracy of the MSX PSC of $\sim2''$ since
we are interested in cross-correlating the MSX data with other catalogues.  We
searched for counterparts of bright stars in the Hipparcos catalogue (see
Perryman et al.\ 1997).  In order to reduce the likelihood of chance
correspondences we used stars in the multi-colour MSX PSC subset with blue
mid-infrared colours (ie $F_8>F_{14}>F_{21}$), and matched against the full
Hipparcos catalogue.  We found 192 matching stars.  Figure 1 shows the results
of this comparison.  Although the bulk of our data do follow the trend found by
Egan et al.\ (1999) in the comparison of MSX PSC data and the MSX Astrometric
Catalog (Egan \& Price 1996), there is a small non-Gaussian tail of sources at
large separation.  We have individually checked the four sources with
separations of 10$''$ or greater.  Three out of the four Hipparcos stars do
appear to be genuine counterparts of the MSX source after inspection of 2MASS
Atlas images, whilst the fourth one is not as clear since it lies in a more
crowded field, and may be a chance coincidence.  The quoted measurement
reliability (which is essentially a gauge of how well a point source matches
the actual data) is also good for all four of these stars (quoted value at
8$\mu$m is 0 or 1, on a scale of 0--9, where 0 is best fit and 9 worst).  The
most obvious features of the 3 clearly discrepant points are that they are
bright ($F_8>100$Jy) and that the MSX image data shows image artefacts.  It is
notable that the centre of the image is offset from the position quoted in the
MSX PSC, apparently because of these artefacts.  These artefacts arise from
cross-talk between detectors for very bright sources in the cross-scan
direction, and detector glints in the in-scan direction 
(see Price et al.\
2001).   Astrometry of sources with some saturated pixels therefore 
may be affected by this.  

Evidence from other individual sources in the catalogues described below also
suggest that very bright point sources may be offset without the measurement
reliability flag being obviously poor.  Furthermore, at least one of the known
massive YSOs in MSX, W75N, shows a similarly large offset from its known
position (almost 30$''$ in error) despite being fainter, and not showing
obvious image defects.  This is more worrying, since it is harder to check for
such errors.  Price et al.\ (2001: see section 4.4) note that a more
sophisticated pointing algorithm was used in creating the MSX images than that
used in the first version of the MSX PSC.  This leaves the possibility that
there are some non-saturated sources in version 1.2 of the MSX PSC with real
astrometric errors.  Overall, these results suggest that there is a small but
non-negligible fraction of MSX PSC sources with very poor astrometry.

The existence of a group with such poor astrometry also led us to check the
accuracy of the quoted fluxes.  We did find some problems in the MSX PSC.  The
most noteable was the existence of a group of sources with detections at more
than one band, but which had a non-detection at one of the key bands for our
multi-colour subset.  As an example, we considered those sources that were
detected at 14 and 21$\mu$m, but not at 8$\mu$m (since these may have very red
colours similar to young stellar objects).  Unfortunately many of these appear
to be erroneously catalogued in the MSX PSC as non-detections at 8$\mu$m. Some
of these are undoubtedly sources in which the detectors were saturated.  A good
example is the case of V1489 Cyg, an oxygen-rich evolved star.  The MSX PSC
gives this as a non-detection at 8$\mu$m, whilst an ISO SWS spectrum shown in
Molster et al.\ (2002) shows the 8$\mu$m flux is $\sim1000$Jy.  The original
MSX image data clearly shows this very bright source, together with
considerable image artefacts due to its large flux density.  In this case all
of the detectors at 8$\mu$m were saturated as noted by Price et al.\ (2001), so
that no {\em useful} data was obtained however.  Only four such cases of hard
saturation are present in the MSX PSC.  In other cases however only some of the
detectors saturate and the MSX PSC flux includes a compensation for this
`missing' flux by fitting the remaining detectors with the known point spread
function.   In some of these cases, the final MSX PSC flux is too inaccurate
to give a reliable measure, and hence is flagged as bad.

In addition, there are also sources detected at 14 and 21$\mu$m but not at
8$\mu$m that lie in extended confused emission regions.  An example of this
kind is given in Table 2, where we present MSX photometry of known MYSOs.  The
MSX PSC fluxes cannot be relied on in these cases.  This is not especially a
problem for the bright sources since we can determine which ones are likely to
be real by high quality detections at the other wavebands.  The same is not
true for sources which are intrinsically fainter, or where the non-detection is
due to a failure to fit a point source in a region of extended emission.  Of
course in the latter case the concept of a point source may be wrong in any
event.

\subsection{Literature Classification of MSX Sources}

We have found that literature searches in order to classify the MSX
sources leads to three different types of object: those with well
established classification, often referred to in many different
papers; those with poorly established classifications, often referred
to either only in one paper or generically as an infra-red source
detected by IRAS; and those with no previous classification at all.

We can illustrate this with our multi-colour subset of 14897 sources
using the SIMBAD database.  If we include all current references from
the literature then 4896 (33\%) have no identification at all within a
10$''$ radius.  We have utilized the SIMBAD object types to classify
the others as shown in Table 1.  If an object has more than one
literature classification we preferentially adopt the classifications
given in the order they appear in Table 1. Over half of the remaining
sources are classified simply as infrared, radio or maser, which does not
actually tell us what type of object it is. Hence, 86\% of these
sources have never been studied in detail before and the vast majority
have never been classified by stellar type. This is further
illustrated if we look at those objects that have been subjected to at
least some study. For this we take all the objects with more than ten
literature references which number 4363. Again three quarters of these
are only known as an infrared source. Less than 6\% of MSX sources with
multi-colour data are well-studied objects of known type.  This
indicates that the mid-infrared emitting population in our galaxy,
even at these relatively bright fluxes, is still mostly unclassified.

Of course in carrying out this simple analysis, we have relied on the
literature classification being correct.  There will be instances in which this
will not be true and the classifications are far from a homogeneous statistical
sample.  However the size of the sample should ensure that the trends we have
found are accurate.

\subsection{Mid-infrared colours of known objects}

As noted above, most of the MSX sources in the multi-colour subset are
previously either unknown or poorly studied.  The main aim of this paper is to
study how well we can use the colours of these sources to determine what type
of object they might be.  First, we need to study what the colours of known
objects are.  We have chosen representative samples of objects from
compilations that are as homogeneous as possible to further analyse the
characteristics of the MSX PSC sources.  These lists form two major classes:
evolved mass losing stars of all kinds and young stars and H~II regions.  The
actual catalogues we used were as follows (approximately in order of increasing
obscuration): the compilation of Alksnis et al.\ (2001) for largely unreddened
carbon stars (these are mostly optically selected); for OH/IR stars we used the
confirmed candidates presented in Chengalur et al.\ (1993); for planetary
nebulae (PN) the catalogue of Acker et al.\ (1992); for Herbig AeBe stars the
compilation of Th\'{e} et al.\ (1994); for compact H~II regions the sources
detected in the radio continuum by Wood \& Churchwell (1989b), Kurtz,
Churchwell \& Wood (1994) and Walsh et al.\ (1998); and methanol maser sources
without detectable radio emission also came from Walsh et al.\ (1998). Finally
for the massive YSOs themselves we have adopted a greatly restricted list
compiled by ourselves to minimize contamination from non-YSOs, and this list is
given in Table 2.   

In addition to these specific catalogues, we also used the classification of
IRAS low resolution spectrograph detections presented by Kwok, Volk \&
Bidelman (1997).  This provides an additional and largely independent means of
identifying different classes of evolved stars.  In particular, their
classification of oxygen rich stellar sources according to the presence of the
9.7$\mu$m silicate feature in emission or absorption provides a clear test of
the dust opacity.  The Kwok et al.\ sample is also a useful source of more
obscured carbon stars missing from the compilation of Alksnis et al.

Figure 2 shows colour-colour plots of all multi-colour MSX PSC sources
for representative regions of the inner ($20^\circ<l<30^\circ$) and
outer galaxy ($100^\circ<l<260^\circ$).  We have chosen the boundaries
of these regions so that there are similar numbers of sources in each
sample.  We have indicated the track of a blackbody of varying
temperature in this and following figures.  In addition, we have also
indicated the magnitude and direction of a typical extinction vector.
The extinction law we have assumed varies as $\lambda^{-1.75}$ for
$\lambda<5\mu$m and uses the data for astronomical silicate from
Draine \& Lee (1984) at longer wavelengths where the silicate feature
dominates in the diffuse ISM and oxygen-rich circumstellar material
(see Draine 1989), although it is not appropriate for carbon-rich
circumstellar dust.  We averaged the appropriate Draine \& Lee data for
the different MSX filters using the filter profiles from Egan et al.\
(1999). The use of the Draine \& Lee (1984) extinction
curve does have one consequence that is clear from Figure 2.  The
$F_{21}/F_{14}$ colour actually gets bluer for increasing extinction
since the 14$\mu$m filter is located in the dip between the 9.7 and
18$\mu$m silicate features.

There are several noteworthy features in Figure 2. The bulk of the sources lie
at the `blue' end of the distribution where the flux ratios are around or below
unity. As we shall see these are mostly evolved stars. For this population
there are clear differences in the colours seen between the inner and outer
galaxy, with the inner galaxy being somewhat redder. Some of this is
undoubtedly due to the effect of the additional line-of-sight extinction
towards the inner galaxy region. However, the direction of the shift in colour
does not agree particularly well with our reddening vectors. An increase in the
extinction at 14$\mu$m to a level similar to that at 12 and 21$\mu$m, or a
decrease in the extinction at 21$\mu$m would give a better agreement.

There is a large population of very red objects that separate out from
the bulk of the stellar sources.  This is clear when the $F_{21}/F_8$
ratio is considered (a constant $F_{21}/F_8$ ratio in the left hand
panel is a diagonal line from top left to lower right).  These are
generally heavily embedded sources in which the central exciting star
has little direct photospheric contribution to the MSX fluxes.

The colours of all classes of objects except those showing
photospheric emission are primarily determined by the opacity and
temperature of the dust surrounding them.  The presence of dust
features can play a large role in some wavebands for some classes of
object.  The most obvious of these are the broad silicate
emission/absorption feature at 9.7$\mu$m (with a secondary broad
feature usually seen in emission near 18$\mu$m), the PAH features at 7.7, 8.6
and 11.3$\mu$m, as well as weaker complexes between 11 and 14$\mu$m and
the SiC feature at 11.3$\mu$m. For nebular sources the strong
forbidden lines such as 8.99$\mu$m [ArIII], 12.8$\mu$m [NeII] and
15.56$\mu$m [NeIII] can also influence the colours.  Additional weaker
molecular absorption lines are also present in the evolved stars.

The role that the dust temperature/opacity and these features play is best
demonstrated by examining the mid-infrared colour-colour diagrams for our
well-studied samples of known types of objects.  Figure 3 shows the equivalent
plots to Figure 2 for our representative samples of `young' sources associated
with massive stars.  The Herbig Ae/Be stars are the least embedded of the young
sources (indeed many were discovered from optical searches).  This is clearly
reflected in them having less red colours, and less excess 21$\mu$m emission
than the other young sources. Our sample of well-known MYSOs is slightly bluer
than the H II regions and methanol sources. This is likely to be because they
are bright, nearby sources whereas the radio selected objects are typically
more distant and suffer more line-of-sight extinction rather than the known
MYSOs being less embedded in their circumstellar material.  All the young
sources are well separated from the bulk of the normal stars in both the
$F_{21}/F_{8}$ and $F_{21}/F_{14}$ ratio, largely due to the fact that they are
embedded in optically thick dust clouds.

A few of these heavily embedded sources have $F_{14}/F_{12}<1$ despite
the fact that $F_{21}>F_{8}$.  For the HII regions this may reflect the
presence of strong 12.8$\mu$m [NeII] emission.  The massive YSO for
which $F_{14}/F_{12}<1$ is S106 IR.  Again there is a relatively
strong [NeII] line present in this object (see the ISO-SWS spectrum
presented in van den Ancker, Tielens \& Wesselius, 2000).  
Although S106 IR has many characteristics of a massive YSO (Drew, Bunn \&
Hoare 1993) it does power a bipolar H~II region.  Most massive YSOs do
not show such ionised gas emission, and hence will tend to have
$F_{14}/F_{12}>1$ reflecting the fact that their continua are red and
mostly featureless in the mid-infrared apart from the silicate
absorption at 9.7$\mu$m (Smith et al.\ 2000). The compact HII regions
have similar colours, but can have stronger forbidden line emission in
virtually any of the filters (e.g. Mart\'{i}n-Hern\'{a}ndez et
al.\ 2002; Giveon et al.\ 2002).
This leads to the slightly larger scatter seen for these objects.

Figure 4 shows the same plots for the samples of evolved sources,
namely carbon stars, OH/IR stars and planetary nebulae.  Here there is
an much greater separation of the various types. The planetary nebulae
have colours similar to the HII regions as we expect since the mid-infrared
emission is dominated by L$\alpha$ heated dust in both cases
(e.g. Hoare 1990; Hoare, Roche \& Glencross 1991). There is a larger
scatter in the PN colours, but their central stars have a much larger
range of effective temperature than those in HII regions, leading to a
wider range in line emission properties. In particular, [NeIII] and
[NeV] in the 14$\mu$m filter can be strong, naturally leading to a
larger scatter in the right hand panel than seen for the HII regions
(de Muizon et al.\ 1990).

The OH/IR stars have moderately red colours and appear to spread out
in the direction of the reddening vector. A better agreement with our
silicate dominated reddening vector is to be expected for these
oxygen-rich stars which produce mostly silicate dust.  To investigate
this further we also looked at the mid-infrared spectral classifications of
evolved stars by Kwok et al.\ (1997).  These results are
shown in Figure 5.  We have only plotted those sources classed types E
(silicate emission at 9.7$\mu$m), A (silicate absorption), C (SiC
emission -- ie carbon stars) and P (PAH emission without noteable
forbidden line emission).  It can be seen that the trends shown by
Figure 4 are confirmed.  The silicate feature in the OH/IR stars
changes from emission to absorption as they cross over the blackbody
line at about $F_{21}/F_{8}$ between 1 and 2 in the right-hand
diagram. The ones with silicate in absorption are more optically thick
and this is consistent with their generally redder $F_{21}/F_{8}$ and
$F_{14}/F_{8}$ ratios. The silicate feature can influence both the 8
and 12$\mu$m fluxes in these objects.  This certainly explains the
behaviour of the OH/IR stars in the right hand panels.  As the
silicate feature goes into absorption, the $F_{14}/F_{12}$ ratio must
rise.  Sevenster (2002) considered the MSX colours of a larger group
of OH masing objects. Her results mostly agree with those here except
that a small group of her sample lie in the region where the PN are
found. Hence, the colours of the OH/IR stars can be understood as a
due to increasing dust opacity as a result of higher dust production
and mass-loss rates.

The carbon stars have the bluest colours of the evolved objects and
lie close to the unreddened blackbody curve. They
can show a strong absorption feature due to various C$_2$H$_2$
transitions near 13.6$\mu$m, as well as weaker absorption in the same
MSX band due to HCN.  The C$_2$H$_2$ features appear stronger in the
more evolved (more obscured) carbon stars detected only in the
infrared (eg Aoki, Tsuji \& Ohnaka, 1999, Volk, Xiong \& Kwok,
2000).  The infrared carbon stars are also more likely to fit an
overall blackbody shape of a few hundred K rather than the combined
effect of the Rayleigh-Jeans tail of the stellar photosphere together
with cool dust emission.  The SiC feature also tends to become weaker
as the obscuration increases (due to self absorption).  The
$F_{14}/F_{12}$ ratio here probably is a closer reflection of the dust
temperature than that of any particular feature, since the effect of SiC
emission at low opacity is probably matched by the effect of the
C$_2$H$_2$ absorption at high opacity.  Most of the sources in both
Figures 4 and 5 are of classical optically selected carbon stars.
Even the most dust enshrouded carbon stars still lie at the extreme red
end of the distribution shown.  We looked at the location of some of
the more extreme dust enshrouded carbon stars studied by
Volk et al.\ (2000) and Aoki et al.\ (1999) using ISO spectroscopy.
Only the two objects in Aoki et al.\ noted as likely post-AGB objects
fall outside the range shown for the carbon stars in Figures 4 and 5
(these lie near the PN population at the cool end of the blackbody
line).

It is now clear that some of the differences between the colour-colour
diagrams for the inner and outer galaxy in Figure 2 are due to changes
in the stellar population present.  There is a significant lack of
carbon stars at the blue end of the right-hand diagram in the inner
galaxy.  Instead there is clearly a large oxygen-rich population
running parallel to the extinction vector and crossing the black body
line near $F_{21}/F_{8}\sim 1-2$.  The increase in the population of
carbon stars in the direction of the anti-centre has been noted before
(eg.\ Jura, Joyce \& Kleinmann 1989), and is usually attributed to
metallicity effects (the lifetime of the carbon star phase increasing
as metallicity decreases).

\section{Near- and Mid-Infrared colours}

\subsection{Sources of near-infrared data}

It is clear from the previous section that although mid-infrared data on their
own allow some separation of object types there is still significant
contamination of the MYSO region with PN in particular.  We require data from
either longer or shorter wavelengths in order to improve this situation.  The
main benefit of using the MSX PSC is the greatly improved spatial resolution
over that provided by IRAS.  This is especially important within the innermost
regions of the Galactic plane. Hence, progress cannot be made with far-infrared
studies until the ASTRO-F satellite makes its far-infrared survey at a
resolution similar to MSX (Shibai 2000). The other available option is to use
near-infrared data from the 2$\mu$m All Sky Survey Point Source Catalogue
(2MASS PSC) (Skrustkie et al 1997).  This was recently completed and the
currently released data covers 47\% of the sky. We have therefore investigated
the use of combined 2MASS and MSX data in the classification of objects and the
identification of young massive stars in particular. Egan, Van Dyk \& Price
(2001) have used combined MSX and 2MASS data in a study of the stellar
populations in the LMC. However, they only used the 8\mic\ MSX data since most
of their sources were undetected in the longer wavelength MSX bands and so they
could not investigate the usage of mid-infrared colour information.  Egan
(1999) also used the first 2MASS incremental data release (a much smaller
sample than considered here) in combination with the MSX PSC to examine object
classification in the Galactic plane.

An immediate problem that we encountered with using 2MASS is that
objects brighter than about 5th magnitude saturate the detectors used.
Since many of the well-known objects, especially the MYSOs, are as
bright as this we have therefore added the compilation of photometry
from the literature by Gezari et al.\ (1999) to the 2MASS PSC data.
Where data exists in both lists we have used the 2MASS data for
consistency, except where the 2MASS data are saturated.  Where data
exists in the literature for sources not currently in the areas
covered by the second incremental data release of 2MASS we have also
used those data.  Lastly, where 2MASS PSC data are flagged as extended
we have used the appropriate data from the 2MASS extended source
catalogue instead.  Only $\sim$20 objects fall into this final
category.

We used a search radius of 10 arcseconds in cross correlating the MSX
source positions and the 2MASS or Gezari catalogues.  This leaves the
possibility that some of our sources are not visible in the near
infrared and that we have picked a neighbouring, but separate, near
infrared counterpart. This is especially true in massive star forming
regions where dense clusters of associated stars are present (e.g
Carpenter et al.\ 1993; Hodapp 1994) and in some cases the true MYSO
may be completely obscured.  We were able to determine some of
the more obvious cases where this was true and remove them.  This was
particularly true for the catalogues presented in Section 2.3, which
represent for the most part well studied objects.  We believe that
$\ls1$\% of our associations are wrong based on this.  Since we are
mostly interested in the statistical properties of the MSX sources, we
will therefore ignore this possibility in what follows.

In objects like G35.2N (Walther et al.\ 1990), GGD27 (Aspin et
al.\ 1994) and IRAS 20126+4104 (Cesaroni et al.\ 1997) where the actual
MYSO is completely obscured at near-infrared wavelengths this procedure will
pick out the nearest star in the cluster or, at 2MASS resolution, a
bright spot in the reflection nebulosity. In these cases, the near-infrared
magnitudes will really be upper limits on the brightness of the true
counterpart. Since, the associated clusters and nebulosities are also
quite faint and red close to the MYSO they still have strong
discriminatory power as we shall see. It is possible that there may be
no 2MASS counterpart at all within 10~arcseconds. With the current
limited sky coverage of 2MASS this is difficult to ascertain
automatically so we only deal with detections and not upper limits here.

We assessed the quality of the available literature data by comparing it with
2MASS where there were overlaps.  Figure 6 shows the comparison of 2MASS and
literature data where both exist (320 objects at $J$, 338 at $H$ and 402 at
$K$). We have only used data for objects with multi-colour MSX data
since these are the objects we are most concerned with in this paper.  There
are several points that must be made before a useful interpretation can be made
of this Figure.  First, much of the older data from the literature was acquired
with single element detectors with a relatively large aperture.  In practice we
discarded all those taken with an aperture larger than 30$''$, and if
more that one measurement was available adopted the one made with the smallest
aperture.  However, there will inevitably be some extended
objects where the larger aperture contains more emission than the 2MASS point
source.  Therefore we would expect deviations from the line in the direction of
fainter 2MASS fluxes.  We may also expect rarer excursions from more recent
imaging data where the aperture reported in the literature is smaller than the
4$''$ resolution of 2MASS, and there is also some evidence for this.  Lastly,
many of the bright sources are evolved stars and
are therefore variable.

Given all of these caveats it is clear that the scatter between 2MASS data and
literature data increases at fainter magnitudes.  The reasons here are likely
to be threefold: first, the aperture effect noted above (though most faint
objects have photometry derived from arrays rather than single element
detectors); second, given an array camera was used, there is a chance that we
are not comparing the same targets, since the literature data may give multiple
point sources within the 2MASS aperture at these faint magnitudes (or the 2MASS
data may contain extended background emission); lastly, there may be an error
in the photometry in either source.  The other obvious feature of Figure 6 is
that there are several objects for which the literature magnitude is fainter
than 5, but 2MASS reports a saturated detection.  We therefore investigated the
source of the 2MASS magnitude for the extreme cases in which a saturated result
is reported for objects with literature magnitudes fainter than 8.  All of
these objects are in areas flagged as being near expected bright stars.  The
area around these sources is masked out in the 2MASS PSC, and a default
saturated magnitude given for the anticipated bright star (note, this also
means that many other sources within these masked strips are missing entirely
from the 2MASS PSC).  We examined the original 2MASS Atlas images and in all
cases the objects were clearly closer to the literature magnitude, and the
original flag marking the source as too bright for reliable 2MASS photometry is
in error.  A smaller number of all the 2MASS sources that are listed as being
saturated are flagged to indicate that they were found to be saturated during
the shortest 2MASS exposure time (the fractions are 0\% at $J$, 11\% at $H$ and
17\% at $K$ of all listed saturated sources).  In these cases the literature
magnitude always agreed with that assessment and so these appear to be
genuinely saturated in 2MASS.

\subsection{Near- and Mid-Infrared Colour-Colour Diagrams and Source
Classification} 

\subsubsection{Sources detected in the MSX PSC at 8$\mu$m only}

The combination of MSX data with near-infrared photometry provides
several potential tests for the nature of the underlying source.  MSX
data on its own is sensitive to the presence of any warm dust.
However, there is little sensitivity to the temperature of any dust
emission and the level of extinction to the source.  Near-infrared
data on the other hand are very sensitive to the extinction. In most
cases the underlying emission will be due to the stellar photosphere
or nebular continuum. Only in the K-band is there likely to be a
contribution from hot dust if present. Hence, the $F_{8}/F_K$ ratio will
increase as we go from simple reddened photospheres to objects with
optically thin warm dust contributing to 8$\mu$m and possibly K and
then onto heavily embedded objects.

Many of the new sources detected by MSX are seen only at 8$\mu$m.  Many of
these are undoubtedly showing only normal photospheric emission, without any
detectable excess due to dust.  As a test of this we selected all sources with
reliable fluxes at 8$\mu$m in the ranges $240<l<250$ and $10<l<11$, but which
were not detected at 14 or 21$\mu$m.  There are 1780 and 2068 objects in the
two respective ranges.  Figure 7 shows the colours of these objects in the
near-infrared and at 8$\mu$m.  The left hand panels can be compared to the
results presented by Egan et al.\ (2001) for the LMC.  The majority of the
sources in the outer galaxy have colours consistent with being cool giant or
main sequence stars.  Within the inner galaxy the results are consistent with a
similar, but much more heavily reddened population.

A small but significant fraction of the objects shown in Figure 7 lie well to
the right of the blackbody line.  These clearly have excess mid-infrared
emission compared to that expected for a standard photosphere.  Most (83\%) of
these mid-infrared excess objects in the outer galaxy have blue near-infrared
colours, and a potential counterpart in the HST Guide Star Catalogue.  Objects
with similar colours are also found in the multi-colour MSX PSC sample.  In the
inner galaxy, most of the sources with excess emission at 8$\mu$m have near
infrared colours close to that expected of normal stars, though the effects of
extinction are more obvious.  We cannot say whether their intrinsic colours are
actually blue as in the outer galaxy.  Most of the sources within the inner
galaxy are unknown, so we cannot be sure of their object classification either.

It seems likely that objects with a mid-infrared excess and blue near infrared
colours are predominantly stars with warm ($\sim200$K), detached, optically
thin dust shells or disks.  Some may be isolated low mass pre-main sequence
stars, which are known to show such an excess.  Other possible candidates for
this mid-infrared excess population include Be stars (cf Oudmaijer et al.\
1992), where the excess is due to emission from the stellar wind rather than
dust, and stars such as Vega with extant debris disks from the star (and
planet) formation process.  Few of the mid-infrared excess sources shown in
Figure 7 are known IRAS sources (most lie near the limit of detection with
MSX).

\subsubsection{The multi-colour MSX sources}

Our primary interest in MSX lies in those red objects with detections at more
than 8$\mu$m however.  We now investigate the additional discrimination from
combining near-infrared and mid-infrared colours.  In Figure 8 we plot the same
inner and outer galaxy samples as were used in Figure 2 on combined near- and
mid-infrared colour-colour diagrams.  Figure 9 shows the same plots as Figure 8
for our selected samples of young sources, and Figure 10 shows the results for
the selected sample of evolved sources.  Note that there are always fewer
points in the right hand panel in these Figures, since they rely heavily on
data from the Gezari et al.\ catalogue, and this often lacks $J$ band data.

Two separate populations appear to be present in the outer galaxy
population in Figure 8(b), one
running along the blackbody curve and the other shifted redwards in
$F_{21}/F_{8}$.  Figure 10 reveals the identity of these
populations. The sequence running along the blackbody curve is due to
carbon stars, whilst the redder one is due to OH/IR stars. In each of
these sequences the reddening (and hence mass-loss rate) is increasing
the further up they appear. This is clear from Figure 11 where we plot
the mid-infrared spectral classifications for evolved stars from Kwok et
al.\ on these combined near- and mid-infrared colour-colour diagrams.  For
the OH/IR stars which have silicate features, this feature is in
emission at smaller $F_{8}/F_K$ ratios and switches to absorption at
the top of the sequence.  The heavily embedded carbon stars discussed
by Aoki et al.\ (1999) and Volk et al.\ (2000) also lie towards the
upper tip of the carbon star sequence, making it likely that the
carbon star sequence is also an extinction sequence.  The evident
extinction sequences show the clear value of adding near-infrared data
to the MSX PSC.

The clearest result of the comparison between inner and outer galaxy shown in
Figure 8 is again a lack of carbon stars within the inner galaxy.  The carbon
star sequence is almost completely absent in Figure 8(a).  There is also
evidence for the effect of greater extinction in the inner galaxy, with
virtually the entire population being shifted to redder $F_{8}/F_K$ ratios.

The separation between the carbon and oxygen rich stars using these
mid- and near-infrared indicators is similar to that proposed by
Epchtein et al.\ (1987) for IRAS and ground-based near-infrared data
(they used the IRAS $F_{25}/F_{12}$ ratio and the ground-based $K-L$
colour).  In both cases we can understand this as being examples of
moderate to high dust obscuration, so that the $K$ band flux reflects
the observed direct light from the star and the 8$\mu$m flux arises
from the warm dust.  It is clear that the near-infrared colours alone
do not follow the same trend however.  Here the stars with silicate in
absorption do not all have redder $F_K/F_J$ colours than those with
silicate in emission.  It is worth noting however that stars with the
silicate feature in absorption do have higher $F_{21}/F_{12}$ ratios
than those with the feature in emission, as we would expect if the
12$\mu$m MSX flux is partially affected by the silicate feature.  It
seems likely that it is the strength of the silicate feature at both
9.7 and 18$\mu$m that gives the actual separation between the two
populations seen.

Comparison of the right-hand panels in Figures 9 and 10 shows that
the majority of PN can now be separated from the H~II regions (and
MYSOs). PN and H~II regions have similar mid-infrared and mid- to
near-infrared colours since they have very similar warm dust
distributions. However, PN mostly have much bluer near-infrared
colours because they are not embedded in a large molecular cloud. This
is certainly the case for evolved PN like NGC~7662 modelled by Hoare
(1990) where their near-infrared colours are dominated by unreddened nebular
continuum. There are some young, compact PN that have hot dust and/or
very small grains/PAHs such as IC 418 (again see Hoare 1990), which
give rise to red near-infrared colours. Other more distant PN will simply
suffer line-of-sight extinction and a few of these can be seen in
Figure 10 overlapping in $F_K/F_J$ with the least reddened H~II regions.

\section{Selection Criteria for Massive Young Stellar Objects}

Our main interest in the MSX PSC is to complete a census of massive
YSOs, as noted in the Introduction.  It is clear from previous surveys
using IRAS data that the primary selection criteria for all massive
YSOs are a rising, largely featureless red continuum between 1 and
100$\mu$m. Therefore, for the MSX PSC data we require
$F_{8}<F_{14}<F_{21}$ as a minimum starting point.  We have started
from the sub-sample of multi-colour MSX PSC data in determining
possible candidates.  6838 (46\%) of the 14897 sources satisfy these
basic colour criteria.  In addition, the massive YSOs in Table 2 also
have $F_{21}/F_{8}>2$.  Therefore we have also imposed these criteria
on the MSX PSC, leaving a sample of 4215 objects.  The rejected
sources are mainly nearby bright stars, evolved stars and a few
planetary nebulae.  If we exclude the innermost 10$^\circ$ of the
plane on either side of the Galactic centre, where source confusion is
highest and distance determination difficult, these numbers are
reduced by a further 30\% to 3071.  We will only deal with this
smaller subset in what follows.

Inspection of Figure 4 shows that are still many evolved stars and PN
that satisfy these selection criteria.  The addition of the near
infrared data does help to considerably reduce the number of such
stars however.  Again, inspection of Figure 9 shows that all the known
massive YSOs satisfy $F_{8}/F_K>5$ and $F_{K}/F_J>2$.  In particular,
these criteria select against most ($\sim2/3$) of the planetary nebulae 
and the
few OH/IR and carbon stars that pass the mid-infrared colour cuts. We
found no significant separation between MYSOs and compact H~II regions
in a pure near-infrared J-H versus H-K colour-colour diagram. Henning
et al.\ (1990) show a separation with compact H~II regions being bluer
in H-K. Unfortunately this probably reflects the photometric technique
employed by Chini, Kr\"{u}gel \& Wargau (1987) who collected the
compact H~II region data by raster scanning for bright sources with a
large beam single-element photometer. This is more than likely to
detect less obscured sources in the vicinity of the true near-infrared
counterpart.

Of the 3071 candidates left after our mid-infrared colour selection, only
736 currently have suitable near-infrared data.  Of these, 472 (64\%)
satisfy the additional colour selection using near-infrared data as
described above.  We expect that this fraction ($\sim2/3$) is
representative of the whole sample: therefore, approximately 2000
sources will have colours in the near- and mid-infrared consistent with
being young massive stars.

There will still be objects other than MYSOs in our colour-selected
sample, most notably compact H~II regions as expected from their very
similar infrared spectral energy distributions.  We carried out a further
literature search on these 472 objects in a similar fashion to that
presented in Section 2.2.  The results of that search are also
presented in Table 1.  Clearly the fraction of evolved or main
sequence stars left in the sample is now greatly reduced compared to
the full sample of multi-colour MSX sources.  In this instance we can
also be reasonably confident that most of the otherwise unknown maser
sources are associated with massive star formation of some kind, and it also
seems likely that the radio sources will predominantly be HII regions.
Discounting those with questionable and infrared
only identifications, there are 221 sources with a
useful classification.  It would appear that 165 (75\%) of these
sources are associated with star formation, confirming that our
selection criteria are reasonable. At least half of these 165 appear
to be H~II regions rather than MYSOs. The maser sources could
be associated with either MYSOs or compact H~II regions.

We also repeated our cross-correlation with the IRAS point source
catalogue for the 3071 candidates. Here 1605 (52\%) of these actually
appear in the IRAS PSC (1466 do not).  Of these 43 have an upper limit
at 60$\mu$m, 560 a upper limit at 100$\mu$m and 408 an upper limit at
both 60 and 100$\mu$m.  Therefore, 2477 (81\%) of our candidates
suffer confusion in the IRAS beam and 1917 (62\%) would
not have been considered in previous searches for MYSOs using IRAS
12, 25 and 60\mic\ colour selection criteria.

We have also examined the distribution in Galactic longitude and
latitude of the MYSO candidates that pass only the mid-infrared colour
selection (the near-infrared data currently has a rather patchy
Galactic distribution so it is better to study only the mid-infrared
selected data in this regard).  Figure 12 shows the results.  The
features present in the longitude distribution are similar to those
found by Wood \& Churchwell (1989a) and mostly correspond to local
complexes and spiral arms.  The high bulge source density within
the inner $\pm10^\circ$ of longitude is clear even for this colour
restricted sample.  The latitude scale height of this sample is
approximately 0.8$^\circ$, similar to the 0.6$^\circ$ found by Wood \&
Churchwell (1989a) for their IRAS selected sample. By comparison, the
scale height of the whole multi-colour MSX sample is 1.6$^\circ$.  The
bulk of the MYSO sample are clearly confined to within $\pm1^\circ$ as
shown by Figure 12(c), as expected if our colour selected sources are
massive stars.  A 0.8$^\circ$ scale height corresponds to 120 pc at
the distance of 8.5 kpc to the galactic centre. This is somewhat
larger than the scale heights of $<$65 pc for local OB stars derived
by Reed (2000) and 30 pc for ultracompact H~II regions by Becker et al.
(1994), probably as a result of the residual contamination of the
colour selected sample by evolved stars.

It is worth briefly considering the effect of the varying sensitivity
limits on our selection criteria.  First, we considered objects with
21$\mu$m detections, that are not detected at 8$\mu$m.  Obviously such
very red sources would be of interest.  Sources that were detected at
14 and 21$\mu$m, but not at 8$\mu$m are mostly artefacts as noted in
Section 2.1.  There are 5431 sources in the MSX PSC with detections
only at 21$\mu$m, however.  Most of these are near the 21$\mu$m
detection limit as well.  A source at the 21$\mu$m detection limit
($F_{21}\sim2-4$Jy) must have an intrinsic $F_{21}/F_{8}$ ratio
greater than $\sim20-40$ to be truly undetected at 8$\mu$m.  This
would certainly place such a source in our region of interest for
finding massive YSOs.  However, given the additional constraint provided by the
limit at 14$\mu$m as well of $\sim1-2$Jy, we can see from Figure 2
that the number of sources expected to fall in the correct part of the
colour-colour space is very small.  We inspected several cases at
random where objects were claimed as detections at 21$\mu$m but not at
shorter wavelengths.  In most instances, where the `detection' is near
the limiting sensitivity, there appears to be no object at any
wavelength in the MSX images.  This appears to be true regardless of
the quality flag present in the MSX PSC.  In a few instances the
`object' is simply part of a larger extended source (and probably was
simply not flagged as a possible point source at the other
wavelengths).  There are certainly many MSX PSC sources which are
actually part of more extended objects rather than point sources at
all flux levels.  By contrast, again by inspection of a random
selection of sources, objects near the detection limit with detections
in more than one band are much more likely to be real.  We believe it
is unlikely therefore that there is a large population of objects
which are genuinely detected at 21$\mu$m, but sufficiently red to be
undetected at any other wavelength.

Of course, there may be a substantial population of objects detected
only at 8$\mu$m which may satisfy our colour criteria. We checked a
random sample of those objects near the 8$\mu$m detection threshold
without detections in other bands as well and they appear to be
reliable. However, we find that the fraction of the 8\mic\ only
sources in Figure 7 that satisfy our two colour cuts involving
near-infrared colours is only 0.15\%. Furthermore from our
multi-colour MSX sample the fraction that pass those cuts and the
further mid-infrared cuts is 40\%. This implies that there are less
than two hundred MYSO candidates detected only
at 8$\mu$m in the full 300~000 source MSX PSC.

This leads naturally to the question of whether we actually expect to detect
all of the MYSOs in the galaxy in our multi-colour MSX sample.  If we assume
the limiting luminosity we wish to detect is 10$^4$\Lsolar, then we can derive
an approximate 21$\mu$m flux for such a source at a distance of 20~kpc.  We
assumed that the bolometric luminosity of these young sources is largely
reprocessed by dust and emerges in the far infrared.  We compared the
far-infrared and MSX fluxes for the sample of known compact HII regions and
MYSOs discussed in Section 2.3.  For simplicity we assumed the far infrared
IRAS flux was given by the relation,
$F_{FIR}=(20.6F_{12}+7.54F_{25}+4.58F_{60}+1.76F_{100})\times10^{-14}$Wm$^{-2}$
(Emerson 1988).  We took the relation between flux and flux density from Cohen,
Hammersley \& Egan (2000) for band E, which is
$F_E=4.041\times10^{-14}F_{21}$Wm$^{-2}$.  Typical $F_{FIR}/F_E$ ratios are in
the range $5-40$.  For our putative 10$^4$\Lsolar\ source on the far side of
the Galaxy, $F_{FIR}\sim10^{-12}$Wm$^{-2}$, and hence
$F_E=2.5\times10^{-14}-2\times10^{-13}$Wm$^{-2}$.  The corresponding $F_{21}$
range is $0.6-5.0$Jy.  Egan et al.\ (1999) estimate the actual completeness
from the source counts as a function of Galactic longitude.  They quote a 50\%
completeness level at 2.5~Jy in the inner two quadrants of the plane, and
3.5~Jy in the outer two quadrants.  This crude analysis suggests that we should
detect at least 50\% of the MYSOs at 10kpc.  Later versions of the MSX PSC are
expected to be a factor of two deeper 
(Egan, private communication), so that we
should actually be able to detect 100\% of the MYSOs within 10kpc eventually,
and all but the extremely reddened sources at 20kpc.  Of course there are
patches of such heavy extinction even in the mid-infrared, as illustrated by
the dark clouds seen at 8\mic\ by MSX (e.g. Carey et al.\ 2000).

Finally, it is worth briefly noting what other observations could be
made to distinguish the MYSOs, compact H~II regions and the remaining
contaminating evolved sources in the colour selected sample. The
Sridharan et al.\ (2002) study shows that single-dish surveys cannot be
relied upon to remove compact H~II regions and in any case their
resolution and positional accuracy is too low to separate compact H~II
regions from nearby MYSOs. High resolution radio continuum
observations can easily distinguish the bright extended free-free
emission from H~II regions and PN from the weak, compact 
stellar wind emission from MYSOs (e.g.\ Hoare
2002). However, there is always the
strong possibility of a MYSO (dominating the mid-infrared flux) being
located very close to a compact H~II region (dominating the radio
flux) such as the case with the BN/KL region behind M42. Missing such
MYSOs would heavily bias the final sample against possible examples of
triggered/sequential star formation. Here, high resolution
ground-based mid-infrared observations are essential since they will
reveal any MYSOs as point sources against the extended emission from
warm dust in the H~II region.

Evolved stars, proto-PN, weak, very compact H~II regions and compact PN will
all be mid-infrared point sources and could be relatively radio quiet. These
will have to be identified by high resolution ground-based near-infrared
imaging. The young sources are usually associated with star clusters,
nebulosity and extinction, which distinguishes them from evolved sources in the
field. The MYSOs often have bipolar or mono-polar reflection nebulae and/or
shocked emission, which appears different to the ionization fronts seen around
compact H~II regions. Near-infrared spectroscopy will be required to identify
any remaining difficult cases.  Kinematic distances will allow an estimate of
the bolometric luminosity and distinguish low-mass and high-mass YSOs.

\section{Conclusions}

We have shown that the combination of the MSX and 2MASS plus literature
near-infrared data is a powerful diagnostic for detecting and classifying many
kinds of embedded source with the Galactic Plane (in agreement with the earlier
results of Egan 1999).  In particular, the combination provides a good
separation between young embedded stars and evolved stars.  Evolved stars can
easily be separated into those which have oxygen-rich dust and those which have
carbon-rich dust, in a similar fashion to that first proposed by Epchtein et
al.\ (1987). The data are also likely to be a valuable tool in detecting normal
stars with excess emission due to dust.  Although most of these are probably
pre-main sequence stars, some may be systems such as Vega, where the dust
emission is believed to arise in the debris disk created during the planet
formation process.

We have developed colour selection criteria that will deliver a sample
of about 2000 objects containing most of the massive young stellar
objects and compact HII regions in our Galaxy. This number is in line
with crude estimates of how many of these objects there should be in
the Galaxy based on the IMF and current star formation rate. A
programme of ground-based follow-up observations of these objects is
already underway to confirm their identity and begin their detailed
characterisation. This will lead to the first large ($\approx$ 1000
objects) and well-selected sample of both MYSOs and compact H~II
regions. Such an unbiased sample will be invaluable in the future
study of massive star formation.

\section{Acknowledgments}

We would like to thank the referee Michael Egan for his helpful comments
on the manuscript and the intricacies of the MSX dataset.
SLL acknowledges the support of PPARC through the award of an Advanced Research
Fellowship.  This publication makes use of data products from the Two Micron
All Sky Survey, which is a joint project of the University of Massachusetts and
the Infrared Processing and Analysis Center/California Institute of Technology,
funded by the National Aeronautics and Space Administration and the National
Science Foundation.  This research has made use of the SIMBAD database,
operated at CDS, Strasbourg, France.

\parindent=0pt

\vspace*{3mm}

\section*{References}
\begin{refs}
\mnref{Acker A., Ochsenbein F., Stenholm B., Tylenda R., Marcout J.,
        Schohn C., 1992, Strasbourg-ESO Catalogue of Galactic 
        Planetary Nebulae, ESO}
\mnref{Alksnis A., Balklavs A., Dzervitis U., Eglitis I., Paupers O., 
	Pundure I.,  2001,  Baltic Astronomy,  10,  1}
\mnref{Aoki W., Tsuji T., Ohnka K., 1999, \aaa, 350, 945}
\mnref{Aspin C., et al., 1994, A\&A, 292, L9}
\mnref{Aumann H.H., 1985, PASP, 97, 885}
\mnref{Becker R.H., White R.L., Helfand D.J., Zoonematkermani S., 
	1994, ApJS 91 347}
\mnref{Beichman C.A., 1987, ARAA, 25, 521}
\mnref{Bunn J.C, Hoare M.G. \& Drew J.E., 1995, MNRAS, 272, 346}
\mnref{Campbell B., Persson S.E., Matthews K., 1989, AJ, 98, 643}
\mnref{Carey S.J., Feldman P.A., Redman R.O., Egan M.P., MacLeod
        J.M., Price S.D., 2000, ApJ, 543, 157}
\mnref{Carpenter J.M., Snell R.L., Schloerb F.P., Skrutskie M.F.,
        1993, ApJ, 407, 657}
\mnref{Caswell J.L., Vaile R.A., Ellingsen S.P., Whiteoak J.B.,
        Norris R.P., 1995, 272, 96}
\mnref{Cesaroni R., Felli M., Testi L., Walmsley C.M., Olmi L., 1997,
        A\&A, 325, 725}
\mnref{Chan S., Henning Th., Schreyer K., 1996, A\&AS, 115, 285}
\mnref{Chengalur J.N., Lewis B.M., Eder J., Terzian Y.,  1993,
	\apjs,  89,  189}
\mnref{Chini R., Kr\"{u}gel E., Wargau W., 1987, A\&A, 181, 378}
\mnref{Codella C., Felli M., Natale V., 1994, A\&A, 284, 233}
\mnref{Cohen, M., Hammersley, P.L., Egan, M.P.,  2000, AJ, 120, 3362}
\mnref{de Muizon J., Cox P., Lequeux J., 1990, A\&AS, 83, 337}
\mnref{Draine B.T., Lee H.M., 1984, \apj, 285, 89}
\mnref{Draine B.T., 1989, in Infrared Spectroscopy in Astronomy,
        Proceedings of the 22nd ESLAB Symposium, p 93}
\mnref{Drew J.E., Bunn J.C, Hoare M.G., 1993, MNRAS, 265, 12}
\mnref{Egan M.P., Price S.D.,  1996,  \aj,  112,  2862}
\mnref{Egan M.P., 1999, Bulletin of the 
	American Astronomical Society, 31, 1507}
\mnref{Egan M.P., et al., 1999, ``The Midcourse Space Experiment Point 
	Source Catalog Version 1.2 Explanatory Guide'', 
     (AFRL-VS-TR-1999-1522) (Springfield, VA: Natl. Tech. Inf. Serv.)}
\mnref{Egan M.P., Van Dyk S.D., Price S.P., 2001, AJ, 122, 1844}
\mnref{Emerson J.P., 1988, in Dupree A. K. \& Lago M. T. V. T.,
        Formation and Evolution of Low Mass Stars, NATO ASI Series C,
        241, Dordrecht, Kluwer, p 193}
\mnref{Epchtein N., Le Bertre T., Lepine J.R.D., Marques Dos Santos
        P., Matsuura O.T., Picazzio E., 1987, A\&AS, 71, 39}
\mnref{Gezari D.Y., Pitts P.S., Schmitz M., Catalog of Infrared
        Observations, Edition 5, On-line Data Catalog}
\mnref{Giveon U., Sternberg A., Lutz D., Feuchtgruber H., Pauldrach A.W.A.,
	 2002, ApJ, 566, 880} 
\mnref{Henning Th., Friedmann C., G\"{u}rtler J., Dorschner J., 1984, Astron. Nactr., 305, 67}
\mnref{Henning Th., Pfau W., Altenhoff W. J., 1990, A\&A, 227, 542}
\mnref{Hoare M.G., 1990. MNRAS, 244, 193}
\mnref{Hoare M.G., Roche P.F. \& Glencross W.M., 1991. MNRAS, 251, 
584}
\mnref{Hoare M.G., 2002, in P.A. Crowther ed., The Earliest
Stages of Massive Star Birth, in ASP Conf. Series, 267, p 137}
\mnref{Hodapp K.-W., 1994, ApJS, 94, 615}
\mnref{Hughes V.A., MacLeod G.C., 1989, AJ, 97, 786}
\mnref{Ishii, M., Nagata, T., Sato, S., Yao, Y., Jiang, Z., Nakaya, H.,
	2001, \aj, 121, 3191}
\mnref{Jura M., 1999, ApJ, 515, 706}
\mnref{Jura M., Joyce R.R., Kleinmann S.G., 1989, ApJ, 336, 924}
\mnref{Kurtz S., Churchwell E., Wood D.O.S.,  1994,  \apjs,  91,  659}
\mnref{Kwok S., Volk K., Bidelman W.P., 1997, ApJS, 112, 557}
\mnref{Lada C.J., 1985, ARA\&A, 23, 267}
\mnref{Lumsden S.L., Puxley P.J., 1996, \mn, 281, 493}
\mnref{Lumsden S.L., Puxley P.J., Hoare M.G., 2001, 328, 419}
\mnref{Mart\'{i}n-Hern\'{a}ndez, N.L., et al., 2002, A\&A, 381, 606}
\mnref{Menten K.M., 1991, ApJ, 380, L75}
\mnref{Molster F.J., Waters L.B.F.M., Tielens A.G.G.M., Barlow M.J.,
	2002, \aaa, 382, 184}
\mnref{Oudmaijer R.D., van der Veen W.E.C.J., Waters L.B.F.M., 
	Trams N.R., Waelkens C., Engelsman E., 1992, \aas, 96, 625}
\mnref{Palla F., Brand J., Comoretto G., Felli M., Cesaroni R., 1991,
        A\&A, 246, 249}
\mnref{Perryman, M.A.C. et al.,  1997,  \aaa,  323,  L49}
\mnref{Plets H., Waelkens C., Oudmaijer R.D., Waters L.B.F.M.,
        1997, A\&A, 323, 513}
\mnref{Porter J.M., Drew J.E., Lumsden S.L., 1998, \aaa, 332, 999}
\mnref{Pottasch S.R., Olling R., Bignell C., Zijlstra A.A., 1988,
        A\&A, 205, 248}
\mnref{Price S.D., 1977, Environmental Reseach Papers, Air Force
        Geophysics Laboratory}
\mnref{Price S.D., Egan M.P., Carey S.J., Mizuno D.R., Kuchar T.A.,
	 2001, AJ, 121, 2819}
\mnref{Prusti T., Adorf H.-M., Meurs E.J.A., 1992, A\&A, 261, 685}
\mnref{Ramesh B., Sridharan T.K., 1997, MNRAS, 284, 1001}
\mnref{Reed, B.C., 2000, AJ, 120, 314}
\mnref{Sevenster, M., 2002, astro-ph/0202183}
\mnref{Shibai H., 2000, AdSpR, 25, 2273}
\mnref{Skrutskie, M., et al., 1997, in The Impact of Large Scale Near-Infrared Sky Survey, ed. F. Garzón et al. (Dordrecht: Kluwer), 25}
\mnref{Smith C.H., Wright C.M., Aitken D.K., Roche P.F.,
	Hough J.H., 2000, MNRAS, 312, 327}
\mnref{Sridharan T.K., Beuther H., Schilke P., Menten K.M., Wyrowski
        F., 2002, ApJ, 566, 931}
\mnref{Th\'{e} P.S., de Winter D., Perez M.R.,  1994,  \aas,  104,  315}
\mnref{Tofani G., Felli M., Taylor G.B., Hunter T.R., A\&AS, 112, 299}
\mnref{Van den Ancker, M.E., Tielens, A.G.G.M., Wesselius, P.R.,  
	2000,  \aaa,  358,  1035}
\mnref{Van der Veen W.E.C.J., Habing H.J., 1988, A\&A, 194, 125}
\mnref{Volk K., Xiong G.-H., Kwok S., 2000, \apj, 530, 408}
\mnref{Walker H.J., Cohen M., 1988, AJ, 95, 1801}
\mnref{Walsh A.J., Burton,M.G., Hyland A.R., Robinson G.,  1998,
	\mn,  301,  640}
\mnref{Walther D.M., Aspin C., McLean I.S., 1990, ApJ, 356, 544}
\mnref{Wood D.O.S., Churchwell E.,  1989a,  \apj,  340,  265}
\mnref{Wood D.O.S., Churchwell E.,  1989b,  \apjs,  69,  831}
\mnref{Wynn-Williams C.G., 1982, ARAA, 20 587}
\end{refs}

\newpage

\onecolumn

\noindent {\bf Table 1:}  The classification from the literature of all 
multi-colour
MSX PSC sources.  We have classed as a YSO any object called 
a YSO, T Tauri star or pre-main sequence star, as an HII region any
object classed as an HII region or young stellar cluster, as a planetary
nebulae any object with a firm identification, as an evolved star any
star with clear mass loss such as a Mira, OH/IR star, carbon star,
or pulsating variable, as a variable star any other kind of variable
star not covered in the evolved star category, as an `other' star any
remaining stellar sources including many emission line stars.
The maser sources are a heterogeneous
collection, given that the transition observed is not considered, so
could pertain to either end of the life of a star.  Radio and infrared
identifications generally imply that the source is not classified but
is known.  ID? refers to objects with debatable classification (often
possible planetary nebulae).

\begin{tabular}{lccccccccccccc}
            & \multicolumn{1}{c}{YSO} & \multicolumn{1}{c}{HII} &
	\multicolumn{1}{c}{PN} & \multicolumn{3}{c}{Star}
	& \multicolumn{1}{c}{Maser} 
	& \multicolumn{1}{c}{Radio} & \multicolumn{1}{c}{IR} 
	& \multicolumn{1}{c}{ID?} & \multicolumn{1}{c}{No ID} & Total \\
 & & & & Evol. & Var. & Other & & & & & \\
 & & & & & & & & & & & \\
All sources & 43 & 200 & 170 & 732 & 331 & 571 & 433 & 295 & 7057 &
	170 & 4896 & 14897 \\ 
 & 0.3\% &1.3\% & 1.1\% & 4.9\% & 2.2\% & 3.8\% & 2.9\% & 2.0\% & 47\%& 1.1\% &33\% & \\
 & & & & & & & & & & & & \\
More than  & 21 & 123 & 123 & 256 & 76 & 247 & 180 & 39 & 3277 & 21
& -- & 4363\\
10 references  &0.5\% & 2.8\% & 2.8\% & 5.9\% & 1.7\% & 5.7\% & 4.1\% & 0.9\% & 75\%
	& 0.5\% & -- & \\
 & & & & & & & & & & & & \\
Subset of & 23 & 53 & 20 & 13 & 1 & 24 & 55 & 32 & 131 & 20 & 100 & 472 \\
YSO candidates & 4.9\% & 11\% & 4.2\% & 2.7\% & 0.2\% & 5.1\% & 12\% & 6.8\% & 28\%
	& 4.2\% & 21\% & \\
 & & & & & & & & & & & & \\
\end{tabular}
\vspace*{10mm}

\noindent {\bf Table 2:} Adopted sample of confirmed massive YSOs.  
Fluxes are taken from the MSX PSC.  Fluxes marked as -- are listed
as non-detections in the MSX PSC, but are clearly the result of problems
of the kind indicated in Section 2.1.

\begin{tabular}{lccrrr}
Source & RA(1950) & Dec(1950) & \multicolumn{1}{c}{$F_{8}$} & \multicolumn{1}{c}{$F_{14}$ }& \multicolumn{1}{c}{$F_{21}$} \\
 & & & \multicolumn{1}{c}{(Jy)} & \multicolumn{1}{c}{(Jy)} 
& \multicolumn{1}{c}{(Jy)} \\
W3 IRS   &  02 21 53.3 & +61 52 21& 183.56 & 851.11 & \multicolumn{1}{c}{--} \\
GL 4029  &  02 57 34.6 & +60 17 23& 6.36&  17.13 &  117.72 \\
GL 437 S &  03 03 32.0 & +58 19 12 &17.08 & 48.64 &  338.12 \\
GL 490   &  03 23 39.2&  +58 36 35 &57.90 & 138.66 & 271.61\\
GL 5180  &  06 05 53.9&  +21 38 57 &5.26 & 13.53 & 78.75 \\
S255 IRS1&  06 09 58.6&  +18 00 13 &101.32  &  156.96  & 263.91\\
GL 961E  &  06 31 59.1 & +04 15 09&45.68  &  84.35  &   315.00 \\
GL 989  &   06 38 24.9&  +09 32 28&92.19  &  160.86  &  262.34\\
NGC6334V 4& 17 16 36.1&  -35 54 47& \multicolumn{1}{c}{--} &   25.76  &   386.49  \\
M8E     &   18 01 49.0 & -24 26 57&78.59  &   138.66  & 230.35\\
W33 A   &   18 11 44.2&  -17 52 58 &15.82  &  50.43  & 168.48 \\
GGD27  &    18 16 13.0 & -20 48 48 &11.33  &  37.24  &  215.46\\
GL 2136&    18 19 37.3 & -13 31 45& 122.61  & 272.40  & 527.54 \\
G35.2N  &   18 55 41.2&  +01 36 27 & 1.99  &  9.79  &  127.55\\
20126+4104& 20 12 41.0 & +41 04 21&0.86  &  4.03  &  53.59 \\
S106 IR &   20 25 33.8 & +37 12 50 & 62.51  &  85.79  & 639.76 \\
GL 2591 &   20 27 35.9&  +40 01 09 &331.46  &  813.72  &  1077.30 \\
W75N   &    20 36 50.0 & +42 26 58 & 15.40  &  60.44  &  476.12\\
Cep A2  &   22 54 19.1&  +61 45 47 &5.05  &  21.33  &  358.41 \\
\end{tabular}

%
%
%
%

\begin{center}
\begin{minipage}{6.5in}{
\psfig{file=Figure1.ps,width=3.5in,angle=270,clip=}}\end{minipage}

\begin{minipage}{\textwidth}{
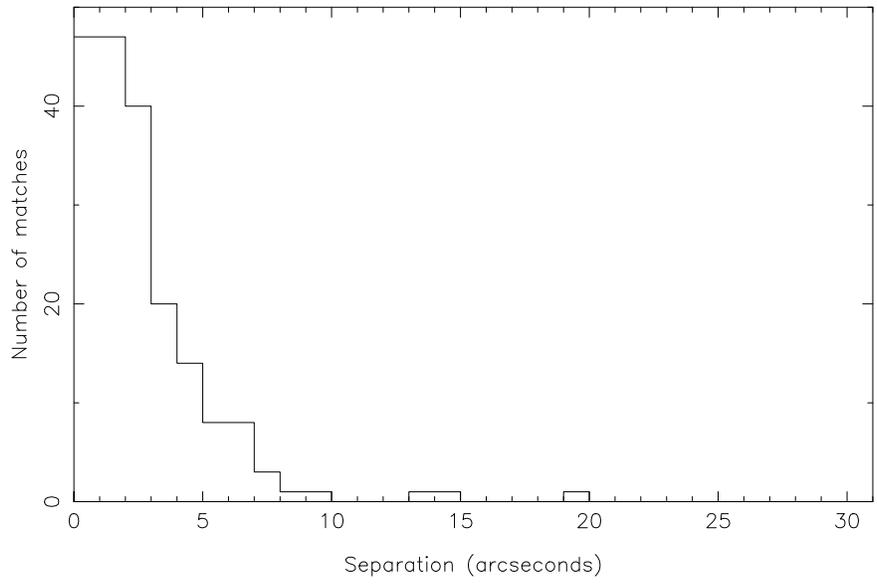
{\bf Figure 1:} The histogram of the separation in quoted position between MSX
PSC and related Hipparcos catalogue entries.  For simplicity, and to avoid
false matches, we have only used MSX PSC sources with overall blue colours.
}\end{minipage}
\end{center}

\begin{center}
(a) \begin{minipage}{6.5in}{
\psfig{file=Figure2a.ps,width=6.5in,angle=0,clip=}}\end{minipage}

(b) \begin{minipage}{6.5in}{
\psfig{file=Figure2b.ps,width=6.5in,angle=0,clip=}}\end{minipage}

\begin{minipage}{\textwidth}{
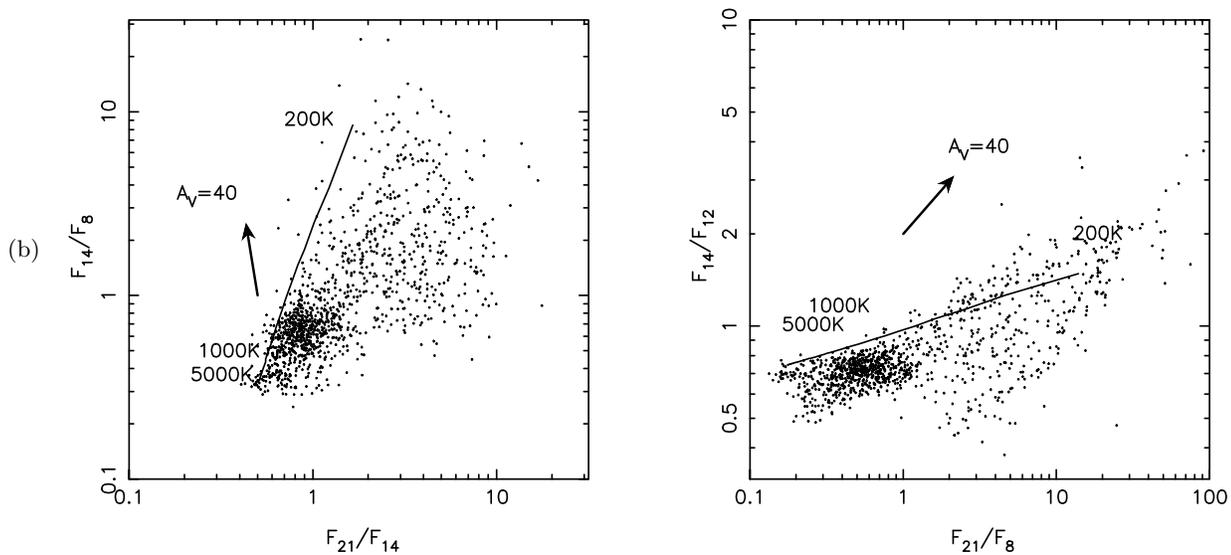
{\bf Figure 2:} Mid-infrared colour-colour plots for (a) a region in the inner
galaxy with $20^\circ<l<30^\circ$ and (b) a region in the outer galaxy with
$100^\circ<l<260^\circ$.  The solid line maps the location of a black body with
temperature between 200K and 5000K.  Also shown is the reddening vector
corresponding to a visual extinction of A$_V=40$.  
}\end{minipage}
\end{center}

\begin{center}
\begin{minipage}{6.5in}{
\psfig{file=Figure3.ps,width=6.5in,angle=0,clip=}}\end{minipage}

\begin{minipage}{\textwidth}{
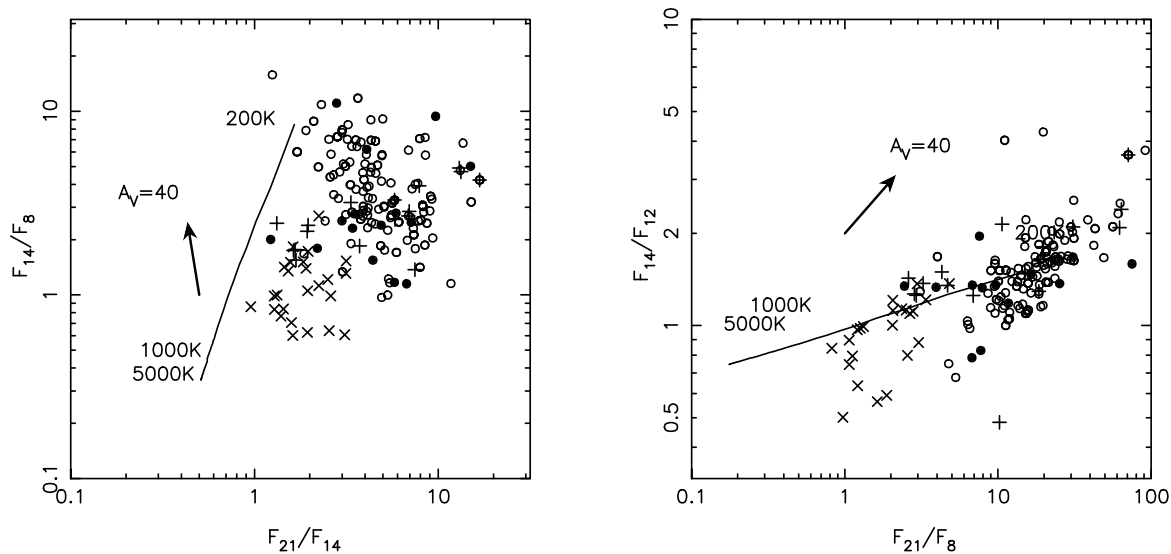
{\bf Figure 3:} Same as for Figure 2 but for samples of known `young' sources
associated with massive stars.  These include Herbig AeBe stars ({\tt x}),
massive YSOs (+), methanol maser sources without radio emission ($\bullet$) and
compact HII regions ($\circ$).
}\end{minipage}
\end{center}

\begin{center}
\begin{minipage}{6.5in}{
\psfig{file=Figure4.ps,width=6.5in,angle=0,clip=}}\end{minipage}

\begin{minipage}{\textwidth}{
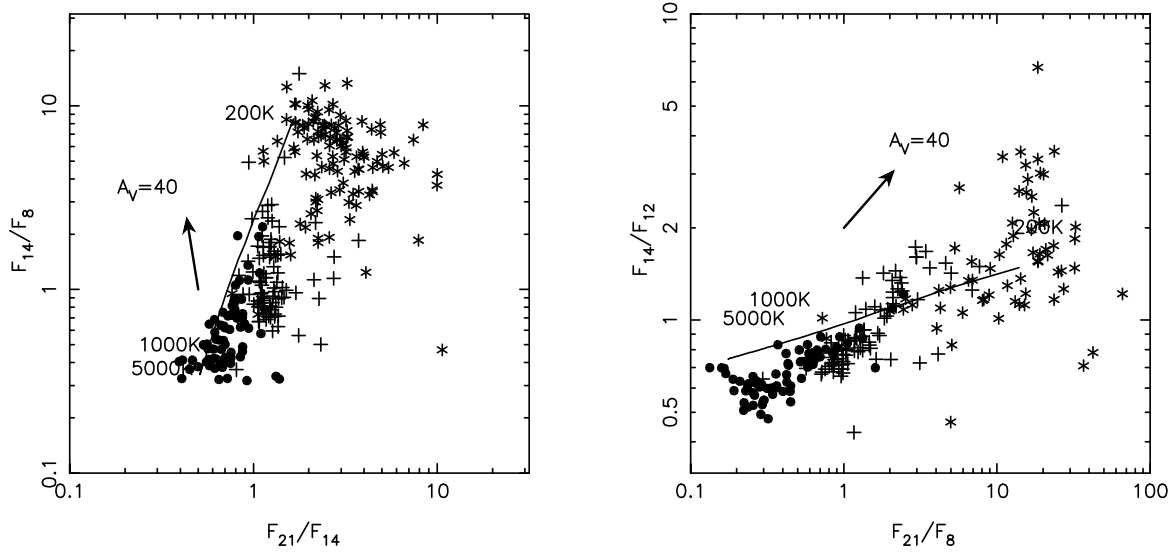
{\bf Figure 4:} Same as for Figure 2 but for samples of known evolved sources.
These include carbon stars ($\bullet$),
OH/IR stars (+) and  planetary nebulae ($\ast$).  
}\end{minipage}
\end{center}

\begin{center}
\begin{minipage}{6.5in}{
\psfig{file=Figure5.ps,width=6.5in,angle=0,clip=}}\end{minipage}

\begin{minipage}{\textwidth}{
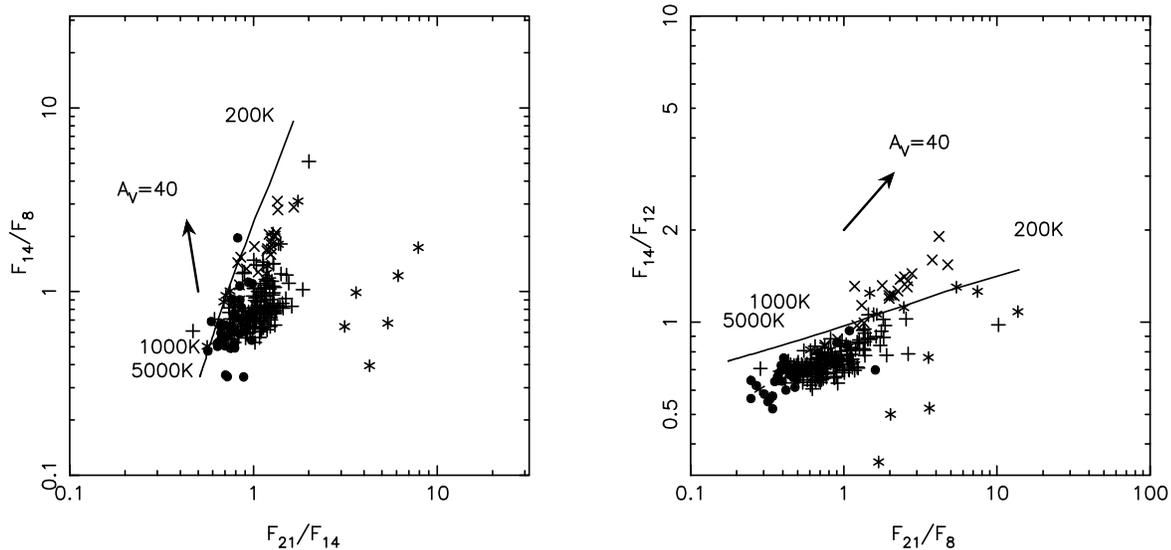
{\bf Figure 5:} Same as for Figure 2 but for sources classified
from IRAS LRS spectra by Kwok et al.\ (1997).  We have only plotted
the carbon stars ($\bullet$), silicate emission objects (+),
silicate absorption objects ({\tt x}) and those classed as having
PAH emission without any ionised gas emission ($\ast$).
For clarity we have only plotted a random selection of one fifth of the
Kwok et al.\ catalogue.  Note in the righthand diagram how the silicate feature moves from
emission to absorption along the OH/IR star sequence in direction of the
plotted extinction vector.  Obscuration in the carbon stars by comparison
tracks the black body curve consistent with a power law dependence
to the extinction.
}\end{minipage}
\end{center}

\begin{center}
\begin{minipage}{6.5in}{
\psfig{file=Figure6.ps,width=6.5in,angle=0,clip=}}\end{minipage}

\begin{minipage}{\textwidth}{
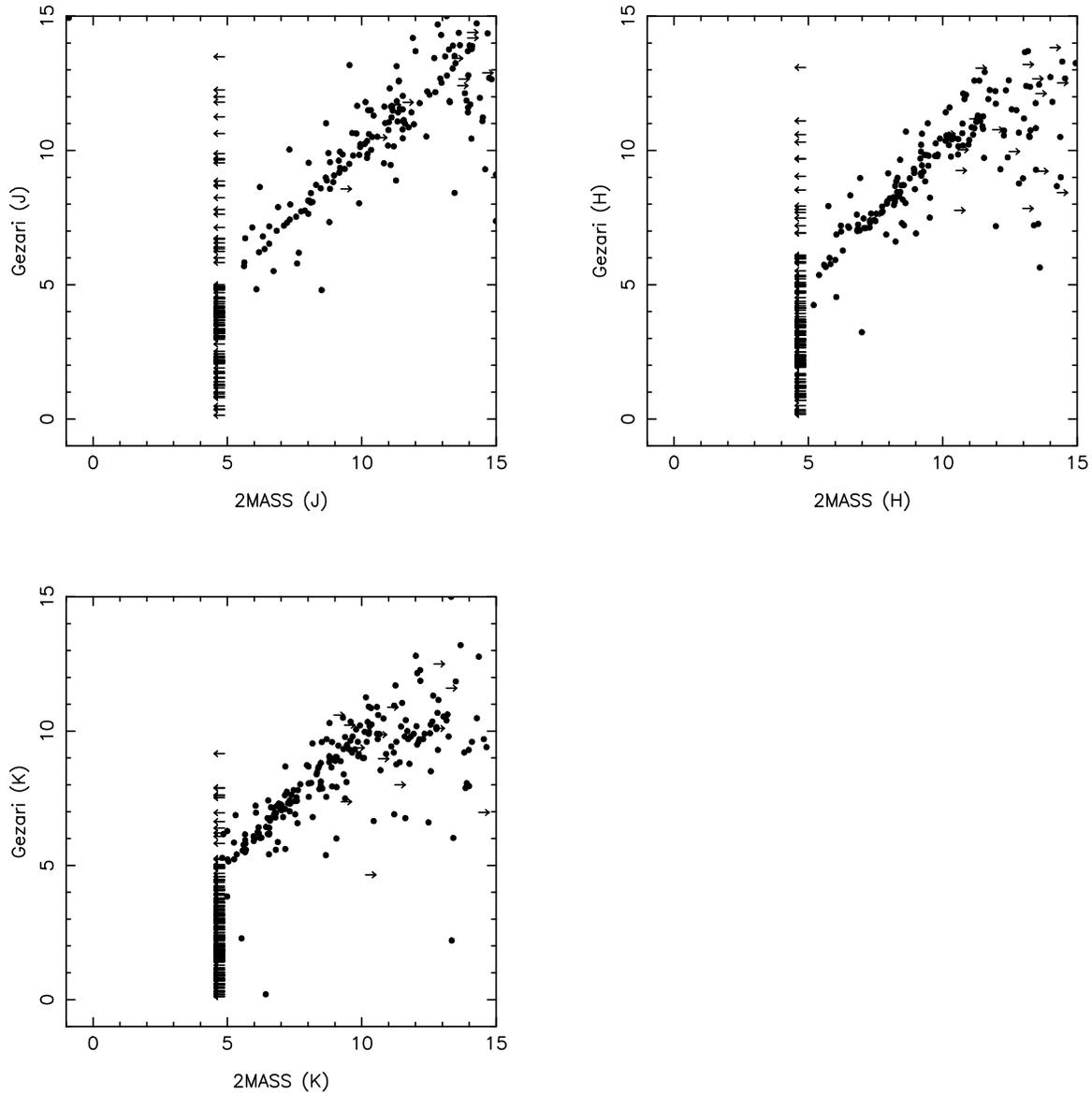
{\bf Figure 6:} Comparison of 2MASS photometry with published data from the
literature for objects where both datasets exists from our multi-colour MSX
PSC sample.  Objects that are noted as saturated in 2MASS are shown
as limits at the nominal saturation point with magnitude$\sim$5.  Lower
limits in the 2MASS PSC are also shown.
}\end{minipage}
\end{center}

\begin{center}
(a)\begin{minipage}{6.5in}{
\psfig{file=Figure7a.ps.cmp,width=6.5in,angle=0,clip=}}\end{minipage}

(b)\begin{minipage}{6.5in}{
\psfig{file=Figure7b.ps.cmp,width=6.5in,angle=0,clip=}}\end{minipage}

\begin{minipage}{\textwidth}{
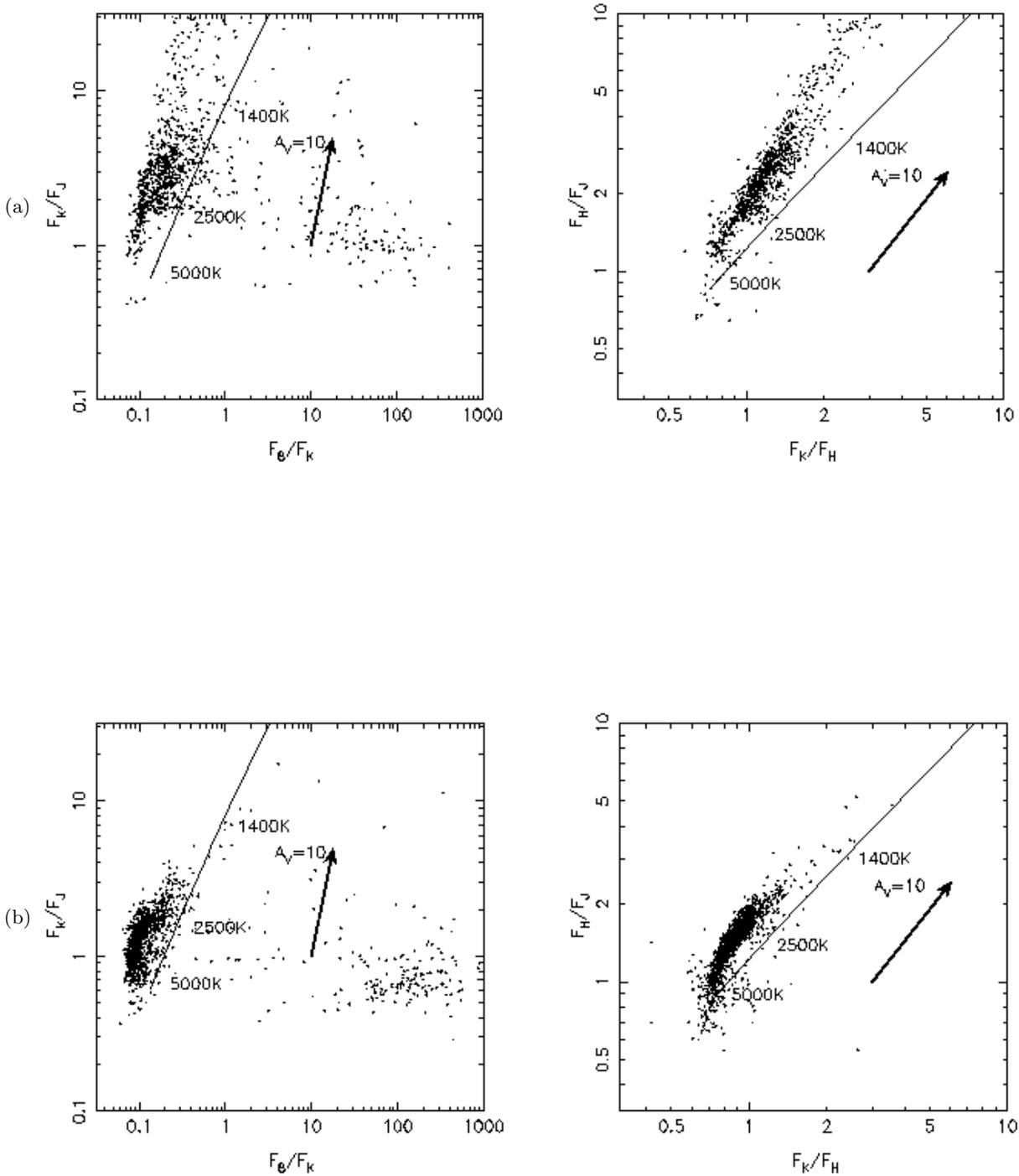
{\bf Figure 7:} Near- and mid-infrared colour-colour plots for sources with
detections only at 8$\mu$m in the MSX PSC.  The data are taken from (a) a
region in the inner galaxy with $10^\circ<l<11^\circ$ and (b) a region in the
outer galaxy with $240^\circ<l<250^\circ$.  The solid line maps the location of
a simple black body.  Also shown is the displacement for a 5000K blackbody due
to a visual extinction of A$_V=10$.  
}\end{minipage}
\end{center}

\begin{center}
(a)\begin{minipage}{6.5in}{
\psfig{file=Figure8a.ps,width=6.5in,angle=0,clip=}}\end{minipage}

(b)\begin{minipage}{6.5in}{
\psfig{file=Figure8b.ps,width=6.5in,angle=0,clip=}}\end{minipage}

\begin{minipage}{\textwidth}{
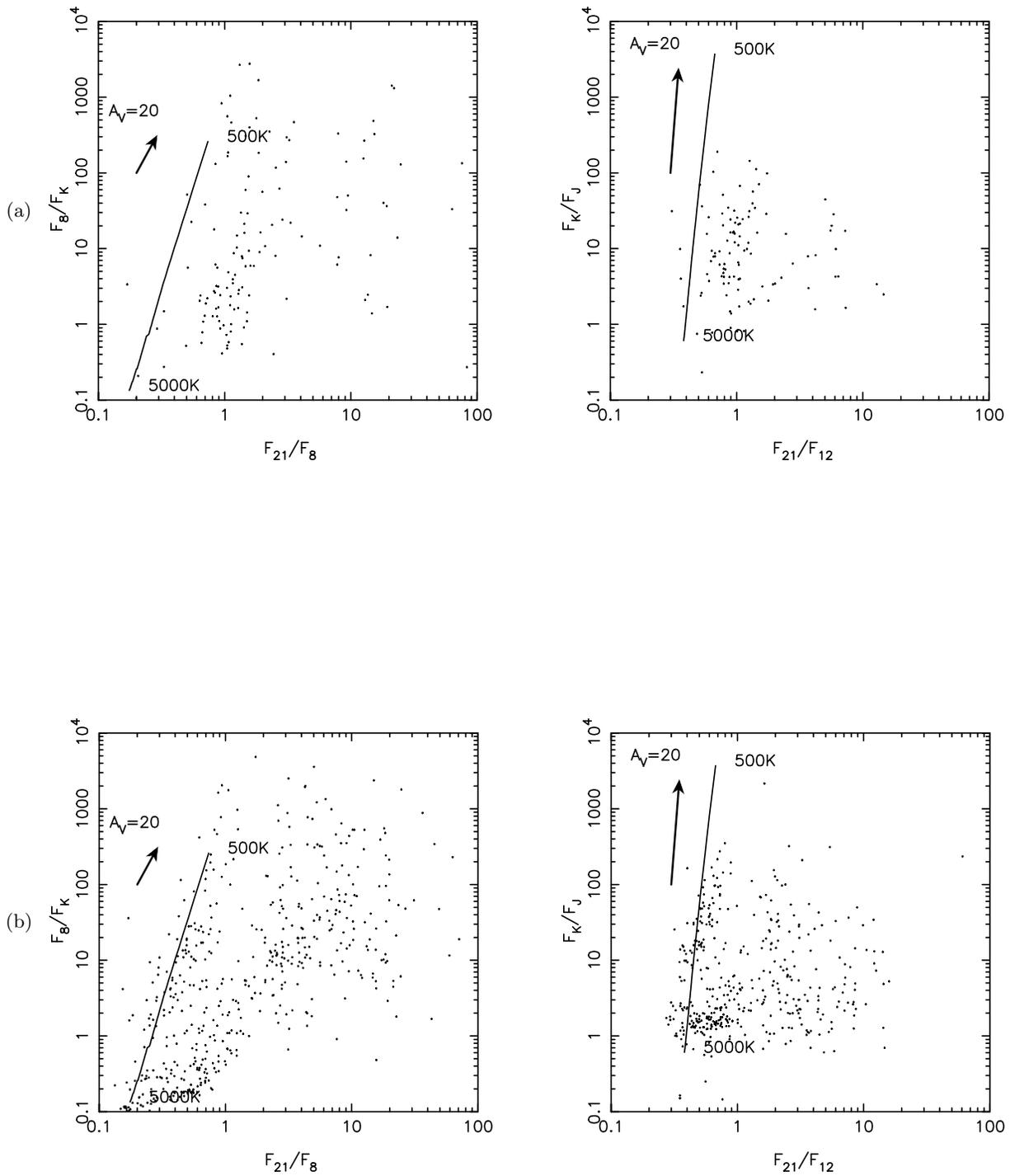
{\bf Figure 8:} Near- and mid-infrared colour-colour plots for the same inner
and outer galaxy samples as shown in Figure 2(a) and (b).  The relative paucity
of K-band detections in Figure 8(a) is actually a result of the patchiness of
the sky coverage of the 2MASS Second Incremental Release.  }\end{minipage}
\end{center}

\begin{center}
\begin{minipage}{6.5in}{
\psfig{file=Figure9.ps,width=6.5in,angle=0,clip=}}\end{minipage}

\begin{minipage}{\textwidth}{
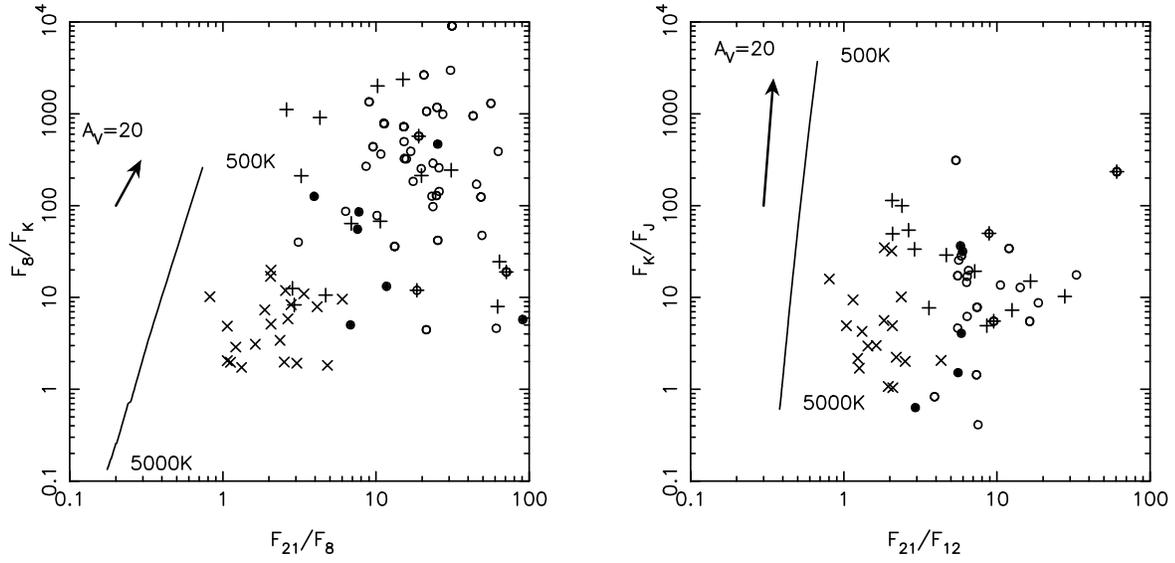
{\bf Figure 9:} As for Figure 8 but for the young sources.  The symbols are
the same as in Figure 3.
}\end{minipage}
\end{center}

\begin{center}
\begin{minipage}{6.5in}{
\psfig{file=Figure10.ps,width=6.5in,angle=0,clip=}}\end{minipage}

\begin{minipage}{\textwidth}{
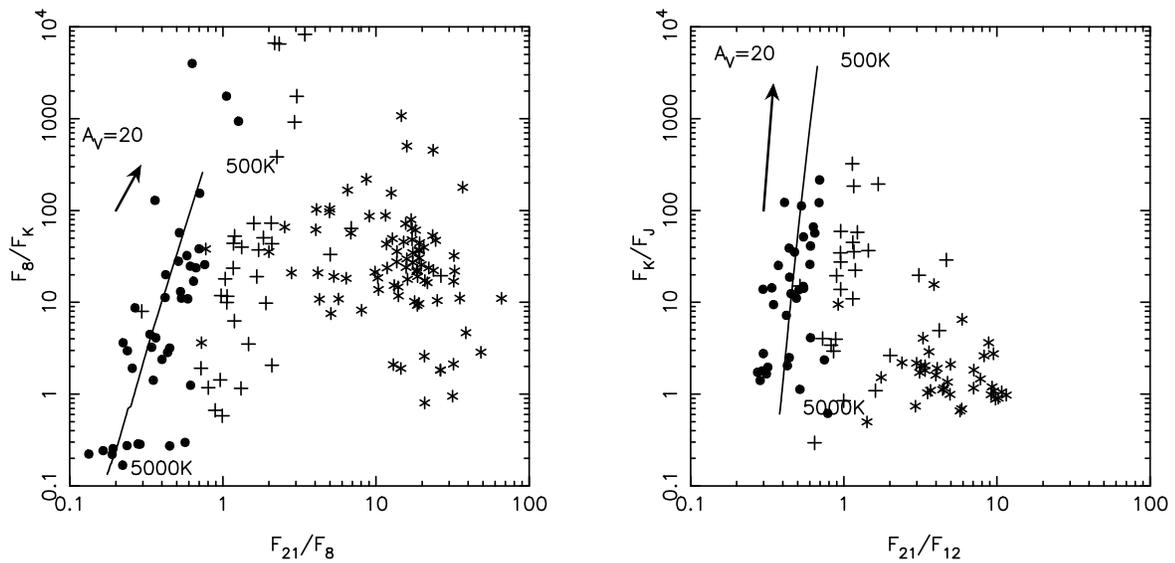
{\bf Figure 10:} As for Figure 8 but for the evolved sources.  The symbols are
the same as in Figure 4.  
}\end{minipage}
\end{center}

\begin{center}
\begin{minipage}{6.5in}{
\psfig{file=Figure11.ps,width=6.5in,angle=0,clip=}}\end{minipage}

\begin{minipage}{\textwidth}{
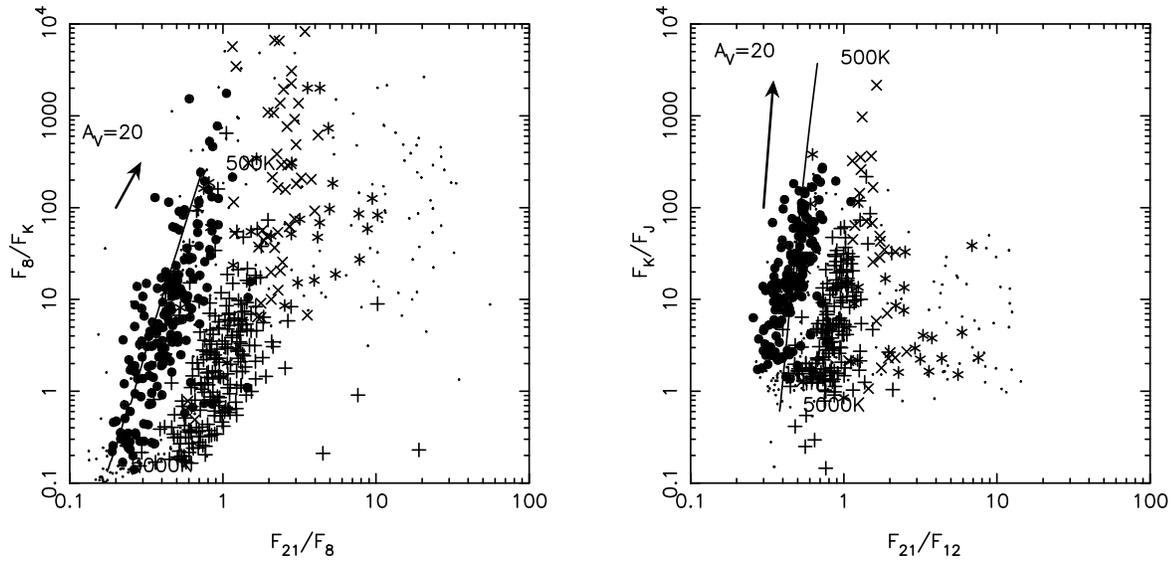
{\bf Figure 11:} As for Figure 8 but for the sources classified by Kwok et
al. The symbols are the same as in Figure 5.  In this case we have plotted all
the points from the Kwok et al.\ catalogue that have near-infrared data.
Objects with classification in Kwok et al.\ (1997)
other than those shown in Figure 5 are shown here
as $\cdot$ for reference.  These include normal stars (near the base of the
blackbody curve) and objects with forbidden line emission (PN and HII regions).
There is a clear separation between oxygen and carbon rich sources in both
panels.
}\end{minipage}
\end{center}

\begin{center}
\begin{minipage}{6.5in}{
\psfig{file=Figure12.ps,width=4.5in,angle=0,clip=}}\end{minipage}

\begin{minipage}{\textwidth}{
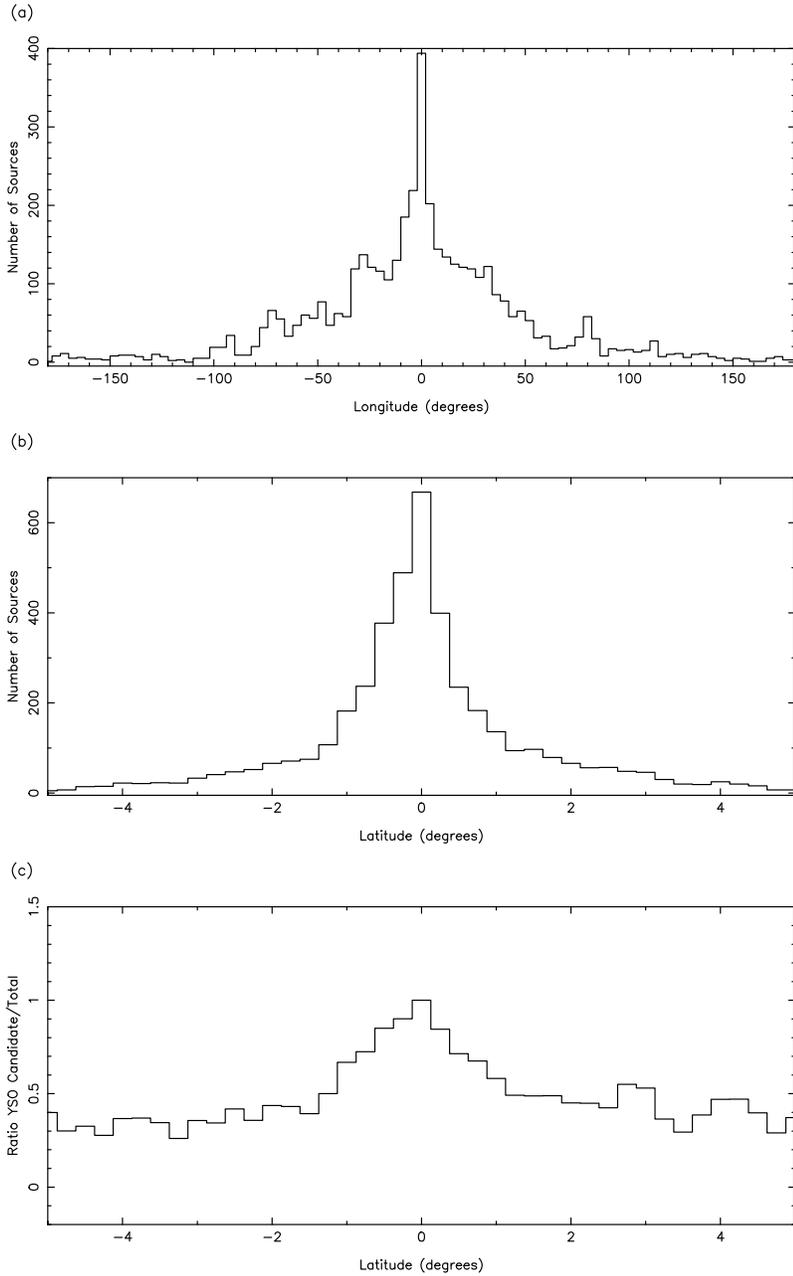
{\bf Figure 12:} Galactic longitude and latitude distribution for the sources
satisfying the mid-infrared colours cuts appropriate to the selection of
MYSOs.  Also shown in (c) is the ratio of the MYSO latitude distribution
with that for the full multi-colour sample, clearly demonstrating the
smaller scale height for the massive star population.
}\end{minipage}
\end{center}

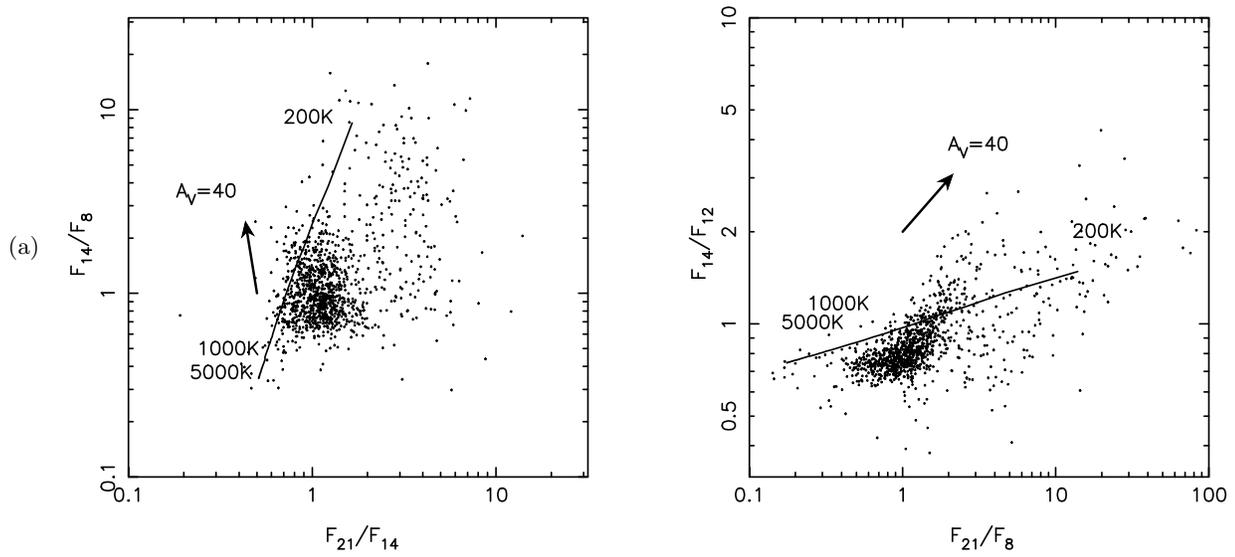
\end{document}